\documentclass{aa}
\usepackage{graphicx}
\usepackage{rotating}
\begin{document}

\newcommand{\as}[2]{$#1''\,\hspace{-1.7mm}.\hspace{.1mm}#2$}
\newcommand{\am}[2]{$#1'\,\hspace{-1.7mm}.\hspace{.0mm}#2$}
\def\approxlt{\lower.2em\hbox{$\buildrel < \over \sim$}}
\def\approxgt{\lower.2em\hbox{$\buildrel > \over \sim$}}
\newcommand{\dgr}{\mbox{$^\circ$}}   
\newcommand{\grd}[2]{\mbox{#1\fdg #2}}
\newcommand{\gsim}{\stackrel{>}{_{\sim}}}
\newcommand{\HI}{\mbox{H\,{\sc i}}}
\newcommand{\HIbf}{\mbox{H\hspace{0.155 em}{\footnotesize \bf I}}}
\newcommand{\HIit}{\mbox{H\hspace{0.155 em}{\footnotesize \it I}}}
\newcommand{\HIsl}{\mbox{H\hspace{0.155 em}{\footnotesize \sl I}}}
\newcommand{\HII}{\mbox{H\,{\sc ii}}}
\newcommand{\IHI}{\mbox{${I}_{HI}$}}
\newcommand{\nan}{Nan\c{c}ay}
\newcommand{\Jykms}{\mbox{Jy~km~s$^{-1}$}}
\newcommand{\kms}{\mbox{km~s$^{-1}$}}
\newcommand{\kmsMpc}{\mbox{ km~s$^{-1}$~Mpc$^{-1}$}}
\def\lir{{\hbox {$L_{IR}$}}}
\def\lco{{\hbox {$L_{CO}$}}}
\def \ls{\hbox{$L_{\odot}$}}
\newcommand{\LB}{\mbox{$L_{B}$}}
\newcommand{\LBnul}{\mbox{$L_{B}^0$}}
\newcommand{\LBsun}{\mbox{$L_{\odot,B}$}}
\newcommand{\lsim}{\stackrel{<}{_{\sim}}}
\newcommand{\LsunK}{\mbox{$L_{\odot, K}$}}
\newcommand{\LsunB}{\mbox{$L_{\odot, B}$}}
\newcommand{\LsunMsun}{\mbox{$L_{\odot}$/${M}_{\odot}$}}
\newcommand{\LK}{\mbox{$L_K$}}
\newcommand{\LKLB}{\mbox{$L_K$/$L_B$}}
\newcommand{\LKLBnul}{\mbox{$L_K$/$L_{B}^0$}}
\newcommand{\LKLsun}{\mbox{$L_{K}$/$L_{\odot,Bol}$}}
\newcommand{\masq}{\mbox{mag~arcsec$^{-2}$}}
\newcommand{\MHI}{\mbox{${M}_{HI}$}}
\newcommand{\MHILB}{\mbox{$M_{HI}/L_B$}}
\newcommand{\MHILBfr}{\mbox{$\frac{{M}_{HI}}{L_{B}}$}}
\newcommand{\MHILK}{\mbox{$M_{HI}/L_K$}}
\newcommand{\KMS}{\mbox{$\frac{km}{s}$}}
\newcommand{\JYKMS}{\mbox{$\frac{Jy km}{s}$}}
\newcommand{\MHILKfr}{\mbox{$\frac{{M}_{HI}}{L_{K}}$}}
\def \ms{\hbox{$M_{\odot}$}}
\newcommand{\Msun}{\mbox{${M}_\odot$}}
\newcommand{\MsunLsun}{\mbox{${M}_{\odot}$/$L_{\odot,Bol}$}}
\newcommand{\MsunLBsun}{\mbox{${M}_{\odot}$/$L_{\odot,B}$}}
\newcommand{\MsunLKsun}{\mbox{${M}_{\odot}$/$L_{\odot,K}$}}
\newcommand{\MT}{\mbox{${M}_{ T}$}}
\newcommand{\MTLBnul}{\mbox{${M}_{T}$/$L_{B}^0$}}
\newcommand{\MTLBsun}{\mbox{${M}_{T}$/$L_{\odot,B}$}}
\newcommand{\tis}[2]{$#1^{s}\,\hspace{-1.7mm}.\hspace{.1mm}#2$}
\newcommand{\Vcor}{\mbox{${V}_{0}$}}
\newcommand{\vhel}{\mbox{$V_{hel}$}}
\newcommand{\VHI}{\mbox{${V}_{HI}$}}
\newcommand{\vrot}{\mbox{$v_{rot}$}}
\def\la{\mathrel{\hbox{\rlap{\hbox{\lower4pt\hbox{$\sim$}}}\hbox{$<$}}}}
\def\ga{\mathrel{\hbox{\rlap{\hbox{\lower4pt\hbox{$\sim$}}}\hbox{$>$}}}}

 \title{A search for Low Surface Brightness galaxies in the near-infrared}

  \subtitle{III. Nan\c{c}ay H{\large \bf I} line observations}

  \author{D. Monnier Ragaigne\inst{1}, 
          W. van Driel\inst{1},
          S.E. Schneider\inst{2},
         C. Balkowski\inst{1}, 
      \and
          T.H. Jarrett\inst{3}
          } 

  \offprints{W. van Driel}

  \institute{Observatoire de Paris, GEPI, CNRS UMR 8111 and Universit\'e  
                Paris 7, 5 place Jules Janssen, F-92195 Meudon Cedex, France \\
             \email{delphine.ragaigne@obspm.fr; wim.vandriel@obspm.fr; 
                    chantal.balkowski@obspm.fr}
        \and
             University of Massachusetts, Astronomy Program, 536 LGRC, Amherst, 
             MA 01003, U.S.A. \\
             \email{schneide@messier.astro.umass.edu}
	\and
             IPAC, Caltech, MS 100-22, 770 South Wilson Ave., Pasadena, 
             CA 91125, U.S.A. \\
             \email{jarrett@ipac.caltech.edu}
             }

\date{\it Received 5/3/2003; accepted 15/5/2003}

  \abstract{ {\rm
A total of 334 Low Surface Brightness galaxies detected in the 2MASS 
all-sky near-infrared survey have been observed in the 21 cm \HI\ line using 
the Nan\c{c}ay telescope. All have a $K_s$-band mean central surface brightness, measured within a 5$''$ radius, 
fainter than 18 \masq\ and a $K_s$-band isophotal radius at the 20 \masq\ level larger than 20$''$. 
We present global \HI\ line parameters for the 171 clearly detected objects and 
the 23 marginal detections, as well as upper limits for the undetected objects.
The 171 clear detections comprise 50 previously uncatalogued objects and 41 objects with a PGC entry only. 
}
  }

 \authorrunning{D. Monnier Ragaigne et al.}
 \titlerunning{A search for LSB galaxies in the near-infrared III.}

 \maketitle

\section{Introduction}  % 1
The present paper is part of a series presenting the results of a 
multi-wavelength (near-infrared, 21-cm \HI\ line and optical) search for Low Surface Brightness (LSB) 
galaxies using the 2MASS all-sky near-infrared survey. 
For further details on the sample and other publications in this series, we refer to 
Monnier Ragaigne et al. (2003a, Paper I). 
\HI\ line observations made at Arecibo of a sample of a further 367 2MASS candidate LSB 
galaxies are presented in Paper II (Monnier Ragaigne et al. 2003b).

\subsection{Low Surface Brightness galaxies}  % 1.1
The brightness of the night sky acts as a filter in the selection of galaxies: when 
convolved with the true population of galaxies it gives the population of galaxies we observe.
In the past few decades, observations of the local universe have shown the
existence of LSB galaxies well below the surface brightness of the average previously 
catalogued galaxies. At present, the LSBs constitute the least well
known fraction of galaxies: their number density and physical properties 
(luminosity, colours, dynamics) are still quite uncertain. This is mainly due to the fundamental 
difficulty in identifying them 
in imaging surveys and in measuring their properties.
In order to further investigate the properties of the LSB 
class of galaxies we selected a large sample of them from the 2MASS database, 
accessing a wavelength domain (the near-infrared) hitherto scarecely explored in the study of LSBs.

There is no unambiguous definition of LSB galaxies, although ones in common use
are based on the mean surface brightness within an isophote or within the half-light radius,
or on the extrapolated central surface brightness of the disc component
alone after carrying out a disc-bulge decomposition.
For 2MASS galaxies, we used the mean $K_s$-band
magnitude within a fixed aperture to identify a sample of galaxies with
relatively low infrared surface brightnesses.
Because galaxies typically have $(B-K)\sim 3.5$--4 (see below) we selected
galaxies in which the central surface brightness within a 5$''$ radius circular aperture
was fainter than 18 \masq in $K_s$. This criterion corresponds
roughly to the disc-component definition of LSBs which have a blue
central surface brightness $\mu_{B_0}>22.0$ \masq. 

LSBs have remarkable properties which distinguish them from high surface 
brightness spirals, notably:
\begin{itemize}
\item LSBs seem to constitute at least 50\% of the total galaxy population in 
number in the local Universe, which has strong implications for the faint end 
slope of the galaxy
luminosity function, on the baryonic matter density and especially on galaxy 
formation scenarios (O'Neil \& Bothun 2000).
\item LSBs discs are among the less evolved objects in the local universe since 
they have a very low star formation rate (van der Hulst et al. 1993; van Zee et
al. 1997; van den Hoek et al. 2000).
\item LSBs are embedded in dark matter halos which are of lower density and 
more extended than the haloes around High Surface Brightness (HSB) galaxies, 
and they are strongly dominated by Dark Matter at all radii (e.g., 
de Blok et al. 1996; McGaugh et al. 2001)
\end{itemize}

The star formation history of LSBs has been the subject of recent debate.
The LSBs best studied in the optical and in the near-infrared are blue
(e.g., Bergvall et al. 1999), indicating a young mean stellar age and/or 
metallicity. 
Morphologically, most studied LSBs have discs, but little spiral structure.
The current massive star formation rates in LSBs are an order of magnitude 
lower than those of HSBs (van Zee et al. 1997); \HI\ observations show that LSBs 
have high gas mass fractions, sometimes exceeding unity 
(Spitzak \& Schneider 1998; McGaugh \& de Blok 1997). All these observations are 
consistent with a 
scenario in which LSBs are relatively unevolved, low mass surface density, 
low metallicity systems with roughly constant star formation rate. However, 
this scenario has difficulty accommodating giant LSBs like Malin 1 
(Bothun et al. 1987).

This study of infrared LSBs was also intended to investigate the possibility
of the existence of a substantial population of red LSBs, like those reported by 
O'Neil et al.~(1997b). Although the \HI\ study of  O'Neil et al.~(2000) indicated that 
some of them did not seem to follow the `standard' Tully-Fisher relation, 
appearing to be severely underluminous for their total mass, 
observations by Chung et al.~(2002) indicate that their rotational properties 
were mismeasured due to confusion with neighbouring galaxies.
An infrared-selected sample should allow us to identify whether there
is a significant population of very red LSBs.

In the present paper results of 21-cm \HI\ line observations made at \nan\ is 
given,  wheras a description of the 2MASS LSB galaxy 
sample selection is given in paper I,  \HI\ observations made at Arecibo are
presented in Paper II, optical BVRI CCD surface photometry of a sub-sample of 
35 galaxies will be presented in paper IV (Monnier Ragaigne et al. 2003d),
an analysis of the full data set will be presented in paper V in the series, and
models of the star formation history of these, and other, samples of LSB galaxies 
are presented in Boissier et al. (2003).

\subsection{The 2MASS all-sky near-infrared survey}  % 1.2
The Two Micron All Sky Survey, 2MASS, has imaged the entire celestial sphere in the 
near-infrared $J$ (1.25$\mu$m), $H$ (1.65$\mu$m) and $K_s$ (2.16$\mu$m) bands from 
two dedicated, identical 1.3-meter telescopes. 
% Its 10-sigma detection level is better than 15.8, 15.1 
% and 14.3 mag at the $J$, $H$, and $K_s$-band, respectively.
The 2MASS data have a 
95\% completeness level in $J$, $H$ and $K_s$ of 15.1, 14.3 and 13.5 mag, 
respectively, for `normal' galaxies (Jarrett et al. 2000; 
for LSB and blue objects the completeness limits are not yet known). 
The Extended Source Catalog (XSC) will consist of more than 1.4 million galaxies 
brighter than 14$^th$ mag at $K_s$ with angular diameters greater than $\sim$6$''$.
The photometry includes accurate PSF-derived measures and a variety of circular and 
elliptical aperture measures, fully characterizing both point-like and
extended objects. The position centroids have an astrometric accuracy better than 
$\sim$\as{0}{5}. In addition to tabular information, 2MASS archives full-resolution 
images for each extended object, enabling detailed comparison with other imaging surveys. 
Initial results for galaxies detected by 2MASS, detailing their properties and detection 
into the Zone of Avoidance (ZoA), are described in several publications 
(Schneider et al. 1997; Jarrett et al. 2000a, 2000b; Hurt et al. 2000). 

Though relatively less deep than some of the dedicated optical imaging surveys
made of LSB galaxies over limited areas of the sky, the 2MASS survey 
allows the detection of LSBs with a central surface brightness in the $K_s$ 
band of $\sim$18-20 \masq, corresponding to about $\sim$22-24 \masq\ in $B$, 
extending over the entire sky. The near-infrared data will be less susceptible 
than optical surveys to the effects of extinction due to dust, both Galactic
and internal to the galaxies.

The selection  of the sample of 2MASS near-infrared selected LSB galaxies
observed in \HI\ at \nan\ is described in Section 2,
the observations and the data reduction are presented in Section 3,
and the results in Section 4. A brief discussion of the results is given in Section 5 and 
the conclusions are presented in Section 6.

\section{Sample selection}  % 2
We have selected 2MASS LSB galaxies using the following two galaxy search 
routines.
For a more complete description of the selection procedures we refer to Paper I.
The selected LSB objects lie outside the ZoA ($|$b$|$$>$10$^{\circ}$).

$\bullet$\, The first is aimed at selecting relatively high signal-to-noise
low central surface brightness (LCSB) galaxies, 
with a mean central $K_s$ surface brightness in the inner 5$''$ 
radius of $K_5$$\geq$18 \masq,
among the extended sources picked out from the
survey data by the standard 2MASS algorithms (Jarrett et al. 2000).
All objects observed in \HI\ at \nan\  have a 20 \masq\ $K_s$-band isophotal 
radius $r_{K_{20}}$ larger than 20$''$, i.e., they belong to our sample of 
``Large'' sources (see Paper I).

$\bullet$\, The second is aimed at finding lower signal-to-noise LSB galaxies 
among those sources which were not recognized as such during the standard
extended source selection described above. 
This requires masking all sources detected by the former method and spatially
smoothing the remaining data. 
These sources are referred to as the ``Faint'' sample, see Paper I.

In order to decide which of these 2MASS sources really are 
LSB galaxies, additional data were used from online 
databases such as NED (NASA Extragalactic Database) 
[http://nedwww.ipac.caltech.edu], LEDA (Lyon-Meudon Extragalactic Database) -- 
recently incorporated in HyperLeda [http://www-obs.univ-lyon1.fr/hypercat/] --
and Aladin of the Centre de Donn\'es astronomiques de Strasbourg (CDS) 
[http://aladin.u-strasbg.fr], and Digital Sky Survey (DSS) images were inspected. 
Using this selection procedure, a total of 3,800 candidate 2MASS LSB 
galaxies were found. 

The source selection for our survey was made with the 2MASS working database 
available in late 1999, when work on it was still in full progress.  
The 229 Large sources we observed at \nan\ contain only 12 that are not found in the 
subsequent, more reliable, 2MASS public data releases (see paper I for details),
6 of which were detected by us in \HI\ (L125O, L191O, L192O, L270O, L668O and L734O)
and 6 that were not (L147N, L328O, L649P, L673O, L761N and L791).
Their names have been put in parentheses in the Tables.
The sample also contains 19 Faint sources, out of the 25 observed at \nan, that are not found in 
subsequent public data releases. This high percentage is not surprising, however, considering 
they were detected by a dedicated LSB galaxy search method and that they were not included 
in the working database. Thirteen of these were detected (see Table 4) and 6 were not 
(see Table 7).

Due to constraints in the useable declination range of the instrument, the area of the sky
observed at Nan\c{c}ay for our survey ranges from  --39$^{\circ}$ to 60$^{\circ}$
in declination, and excludes the declination range we observed at Arecibo, i.e.
0$^{\circ}$ to +39$^{\circ}$. 
Within the chosen area we selected for observation, in order of decreasing priority: 
(1) all objects with known velocity, and
(2) objects without known velocity, whether previously catalogued or not.

\section{Observations and data reduction }  % 3
\subsection{Observations} % 3.1
The \nan\ decimetric radio telescope, a meridian transit-type instrument
of the Kraus/Ohio State design, consists of a fixed spherical mirror (300~m long
and 35~m high), a tiltable flat mirror (200$\times$40~m), and a focal carriage 
moving along a curved rail track. Sources on the celestial equator can be tracked 
for about 60 minutes. The telescope's collecting area is about 7000~m$^{2}$ 
(equivalent to a 94-m diameter parabolic dish). 
Due to the E-W elongated shape of the mirrors, some of the instrument's
characteristics depend on the declination at which one observes. At 21-cm 
wavelength the telescope's half-power beam width (HPBW) is \am{3}{5}~in right 
ascension, independent of  declination, while in the North-South direction it is 23$'$ 
for declinations up to $\sim$20$^{\circ}$, rising to 25$'$ at $\delta$=
40$^{\circ}$ and to 29$'$ at $\delta$= 60$^{\circ}$, the northern limit of the survey 
(see also Matthews \& van Driel 2000). The instrument's effective collecting area 
and, consequently, its gain, follow the same geometric effect, decreasing 
correspondingly with declination.
All observations for our project were made after a major renovation of the
instrument's focal system (e.g., van Driel et al. 1997), which resulted in 
a typical system temperature of 35~K.

The observations were made in the period July 2000-June 2002, 
using a total of about 1500 hours of telescope time.
We obtained our observations in total power (position-switching) mode
using consecutive pairs of 40 seconds ON and 40 seconds 
off-source integrations. Off-source integrations were taken at a position
about 20$'$~E of the target position.
Different autocorrelator modes were used for the observation of sources
with previously known radial velocities and for velocity searches of
objects of unknown redshift.

For objects of known redshift, the autocorrelator was divided into 1 pair
of cross-polarized receiver banks, each with 4096 channels and a 25~MHz 
bandpass, resulting in a channel spacing of 1.3~\kms. The centre 
frequencies of the 2 banks were tuned to the redshifted \HI\ line
frequency of the target source.
These spectra were boxcar smoothed to a channel separation of 17.1~\kms\
during the data reduction in order to increase signal-to-noise.  

For objects without a known redshift, the autocorrelator was divided into
1 pair of cross-polarized receiver banks, each with 4096 channels and a 50~MHz 
bandpass, resulting in a channel spacing of 2.6~\kms. The centre 
frequencies of the 2 banks were tuned to 5600 \kms, 
for a velocity search in the $\sim$-500 to 10,500 \kms\ range (hereafter
referred to as a low-velocity search) and to 15,000 \kms,
for a velocity search in the $\sim$9500 to 20,500 \kms\ range for the
`high-velocity searches'. These spectra were boxcar smoothed to a channel 
separation of 15.7~\kms\ during the data reduction.  

Flux calibration, i.e., the conversion of observed system temperatures to
flux densities in mJy, is determined for the \nan\ telescope through
regular measurements of a cold load calibrator and periodic monitoring of strong
continuum sources by the \nan\ staff. Standard calibration procedures include
correction for the abovementioned declination-dependent gain variations of the 
telescope (e.g., Fouqu\'e et al. 1990). We observed a number of calibrator galaxies 
throughout our observing runs, see Section 5.

\subsection{Data reduction}  % 3.2
The first steps in the data reduction were made using software developed by the 
\nan\ staff (NAPS, SIR program packages).
With this software we averaged the two receiver polarizations, and applied a 
declination-dependent conversion factor to convert from  units of system 
temperature to flux density in mJy. 

Further data analysis was performed using Supermongo routines developed by one
of us (SES). With these we subtracted baselines (generally third order polynomials
were fitted), excluding those velocity ranges with \HI\ line emission or
radio frequency interference (RFI).
Once the baselines were subtracted, the radial velocities were corrected to the
heliocentric system, according to the conventional optical definition
($V=c$($\lambda$-$\lambda_0$)/$\lambda_0$) and the central line velocity,
line widths at, respectively, the 50\% and 20\% level of peak maximum (Lewis 1983), the
integrated flux of the \HI\ profiles, as well as the rms noise of the
spectra were determined. All data were boxcar smoothed to a velocity resolution
of 15.7 \kms\ (velocity search) and  17.1 \kms\ (known velocity) for further analysis.

\subsubsection{RFI mitigation}  % 3.3
As a consequence of their high sensitivity, radio astronomy telescopes are vulnerable
to radio frequency interference (RFI), with signal strengths usually greatly exceeding 
those of the weak celestial radio sources being observed. Broad-band RFI raises the
noise level of the observations, while narrow-band RFI may mimic spectral lines, like
the \HI\ lines from galaxies that were searched for in the present study.

In a regulatory sense, for 21cm line observations, it should be noted that in the 
ITU-R Radio Regulations (2001), with which all users of the radio spectrum are 
obliged to comply, astronomical \HI\ line observations 
are only protected from ``all emissions'' out to a redshift of about 4300 \kms, 
while for observations out to $\sim$19,000 \kms\ ``[national frequency] 
administrations are urged to take all practicable steps to protect the radio astronomy 
service from harmful interference''. These provisions for protection from RFI  
clearly cannot guarantee a completely interference-free environment for
the kind of survey we performed.

At present no universal RFI mitigation method exists for radio astronomy observations 
(e.g., Fridman \& Baan 2001).
In order to minimize the effect of RFI in our observations we used an off-line RFI 
mitigation program, which is part of the standard \nan\ NAPS software package.
The different steps applied to the individual 40 second integration spectra within a
cycle of consecutive spectra of a particular object are as follows:

First, the average of all OFF spectra in the cycle is subtracted from each 
individual ON spectrum, and then for each resulting individual spectrum the 
average signal strength and its rms variation is determined, after rejection of 25\% 
of the highest and 25\% of the lowest signal strengths. 
In each individual spectrum the channels with signal strengths
deviating by more than $\pm$N$\sigma$ from the average are flagged, and their 
signal strengths replaced by a linear interpolation between neighbouring clean 
channels. The cut-off level was set to 10$\sigma$. 

Then the average of all cleaned ON spectra in the cycle is subtracted from each 
individual OFF spectrum. In analogy with the first step, in each resulting individual
spectrum the channels with signal strengths exceeding $\pm$N$\sigma$ from the 
average are replaced by linear interpolation. This results in cleaned individual 
OFF spectra.

Finally, the first step is repeated on the individual ON spectra, but this time 
using the cleaned OFF spectrum resulting from the second step.

% This method is simple, but quite effective for our purpose, as can be seen from a 
% comparison between the two spectra shown in Figure 1, without and with the 
% application of the algorithm. {\bf To be done.....}

Nevertheless, no RFI mitigation technique can simply eliminate all interference (certainly
not without the risk of removing part of the line emission that is being searched for), 
and the detections in the following radial velocity ranges, where the strongest RFI signals 
occur, should be considered with due care: RFI occurred frequently around 4200, 8250, 12,500 and 
17,200 \kms, regularly around 9000 and 11,700 \kms, and occasionally around 2000, 9400 and
10,500 \kms.

\section{Results}  % 4
Our global Nan\c{c}ay \HI\ data for the observed sample, together with their global 
near-infrared and optical data, are listed in Table 3 for detected objects, 
in Table 4 for marginal detections and in Table 5 for undetected objects. 
The \HI\ spectra of the detected galaxies are shown in Figure~1; for information on 
cases where 2 line profiles were detected in the same spectrum, which are designated 
by `a' and `b' after the 2MASS source name, see Sect. 4.5.
For marginal detections, the \HI\ spectra are shown in Figure~2.

A description of all parameters listed in the Tables is given below in Sections 
4.1-4.4. The near-infrared data listed were taken from the 2MASS catalogue
and the optical data were taken from the online NED and LEDA databases, as 
indicated.

The distribution of the integrated line fluxes and FWHM's of the clear and marginal detections 
is shown in Figure 3. Also plotted in this Figure is a straight line
indicating the 3 $\sigma$ detection limit for a 250 \kms\ wide, flat-topped spectral line,
based on the average rms noise level of the data. 
The data quality and the rms noise in the observations is not very uniform
in general, not even for the spectra in which detections were made, resulting in
the dispersion among the weakest clear and marginal detections.

\subsection{Names, positions and distances}  % 4.1
$\bullet$\, Number: we have divided the selected 2MASS sources according to 
two criteria: size and algorithm. This division is indicated
by two characters in the galaxy designations which we will use throughout this series: 
`L' indicates that all objects we observed at \nan\ are ``Large'' object with an isophotal 
$K_s$-band radius $r_{K20}$ (see Sect. 4.2)  larger than 20$''$ (note that no ``Small'' sources, with a radius between 
10$''$ and 20$''$, were observed at \nan) and an 'F' a galaxy selected using the LCSB source 
processor, while following the source number an `O' indicates  a previously catalogued object, 
a `P' one with a PGC entry only and an `N' a previously uncatalogued one.  

\noindent
$\bullet$\, Identifications: for each of the 2MASS sources we queried the NED 
and LEDA databases for a cross-identification of its position with other 
catalogues. 
For previously catalogued objects we list the most commonly used identification 
besides the 2MASS identification;

\noindent
$\bullet$\, Positions: the listed right ascensions and declinations are the 
catalogued 2MASS source centre positions, for epoch J2000.0. These were used 
as the pointing centres for the \HI\ observations; 

\noindent
$\bullet$\, Distances: for each detected galaxy, a distance $D$ was calculated 
using radial \HI\ velocities corrected to the Galactic Standard of Rest,
following the procedure given in the RC3 (and used in LEDA),
and assuming a Hubble constant of $H_0$=75~km~s$^{-1}$~Mpc$^{-1}$.

\subsection{Near-infrared data}  % 4.2
\noindent
$\bullet$\, $K_{20}$ is the total magnitude measured within the $r_{K_{20}}$ isophotal aperture (see below);

\noindent
$\bullet$\, $J-K$ is the $(J-K)_{20}$ near-infrared colour, based on magnitudes measured
in the $J$ and $K_s$ bands within their respective isophotal semi-major axes at the 
20 \masq\ level;

% \noindent
%$\bullet$\, $H-K$ is a near-infrared colour of the source, based on magnitudes measured
%in the $H$ and $K_s$ bands within their respective isophotal semi-major axes at the 
%20 \masq\ level;

\noindent
$\bullet$\, $\mu_{K5}$ is the mean central surface brightness (in \masq) 
measured within a radius of 5 arcsec around the source's centre;

\noindent
$\bullet$\, $b/a$ is the infrared axis ratio determined from an ellipse fit to the co-addition of the 
$J$-, $H$-, and $K_s$-band images. The fit is carried out
at the 3-$\sigma$ isophotal level relative to the background noise in each
image. The 2MASS F (LCSB sources) sample was not measured in this way because of
the low signal-to-noise levels of the emission;

\noindent
$\bullet$\, $r_{K_{20}}$ is the fiducial aperture (in arcsec) in $K$ band. 
Essentially, it is the aperture size for a surface brightness of 20 \masq\ in $K$ 
band;

\noindent
$\bullet$\, \LK\ is the absolute magnitude in the $K$ band (in \LsunK), calculated
using an absolute solar $K$-band magnitude of 3.33 (Allen 1973 ).

\subsection{Optical data}  % 4.3
$\bullet$\, Type is the morphological type, as listed in NED;
% {\it To be done: - change from LEDA to NED values in the Tables};

\noindent
$\bullet$\, $V_{opt}$ is the mean heliocentric radial velocity (in \kms), as listed in LEDA;

\noindent
$\bullet$\, $D_{25}$ is the isophotal $B$-band diameter (in units of  arcmin) 
measured at a surface brightness level of 25 \masq, as listed in LEDA;

\noindent
$\bullet$\, $B_{T_{c}}$ is the total apparent $B$-band magnitude reduced to the
RC3 system (de Vaucouleurs et al. 1991) and corrected for galactic extinction, 
inclination and redshift effects (see Paturel et al. 1997, and references 
therein), 
as listed in LEDA;

\noindent
$\bullet$\, $\mu_{B_{25}}$ is the mean $B$-band surface brightness (in \masq)
within the 25 \masq\ isophote, as listed in LEDA;

\noindent
$\bullet$\, \LB\ is the absolute magnitude in the B band (in \LsunB), calculated
using the $B_{T_{c}}$ magnitude and an absolute solar magnitude in the B band of 
5.48 (Allen 1973), as listed in LEDA.

\subsection{\HIit\ data}  % 4.4
The global \HI\ line parameters are directly measured values; 
no corrections have been applied to them for, e.g., instrumental 
resolution or cosmological stretching (e.g., Matthews et al. 2001)

\noindent
$\bullet$\, rms is the rms noise level in a spectrum (in mJy);

\noindent
$\bullet$\, \IHI\ is the integrated line flux (in \Jykms).
The upper limits listed are 3$\sigma$ values for flat-topped profiles 
with a width of 250 \kms, a representative value for the galaxies detected;

\noindent
$\bullet$\, \VHI\ is the heliocentric central radial velocity of a line profile 
(in \kms), in the optical convention. We estimated the uncertainty, 
$\sigma_{V_{HI}}$  (in \kms), in \VHI\ following Fouqu\'e et al. (1990), as 
\begin{equation}
\sigma_{V_{HI}} = 4 R^{0.5}P_{W}^{0.5}X^{-1}
\end{equation}
where $R$ is the instrumental resolution 
($\sim$17 \kms, see Section 3.1), 
$P_{W}$=($W_{20}$--$W_{50}$)/2 (in \kms) and $X$ is the signal-to-noise ratio of 
a spectrum, which we defined as the ratio of the peak flux density and the rms 
noise;

\noindent
$\bullet$\, $V_0$ is the \HI\ radial velocity (in \kms) corrected to the 
Galactic Standard of Rest, following the RC3;

\noindent
$\bullet$\, $W_{50}$ and $W_{20}$ are the velocity widths (in km/s) at 
50\% and 20\% of peak maximum (km/s). According to Fouqu\'e et al. (1990),
the uncertainty in the line widths is 2$\sigma_{V_{HI}}$ (see above) 
for W$_{50}$ and 3$\sigma_{V_{HI}}$ for W$_{20}$.
 
\noindent
$\bullet$\, \MHI\ is the total \HI\ mass (in \Msun), 
\MHI = 2.356 $10^5 D^2$ \IHI;

\noindent
$\bullet$\, \MHILB\ is the ratio of the total \HI\ mass to the $B$-band 
luminosity (in \MsunLBsun);

\noindent
$\bullet$\, \MHILK\ is the ratio of the total \HI\ mass to the $K_s$-band 
luminosity (in \MsunLKsun).

\subsection{Notes on individual galaxies}  % 4.6
In order to identify possible confused sources in our \HI\ observations,
we first inspected DSS images over a region of 12$'$$\times$36$'$ 
($\alpha$$\times$$\delta$) centred on the 2MASS position of each clearly or
marginally detected source.
In case galaxies were noted that might give rise to confusion with the
\HI\ profile of the target galaxy, we queried the NED and LEDA databases for
data on the objects. The data listed here were preferentially taken from the
mean values listed in LEDA. 
The global \HI\ line parameters of galaxies reported 
as detected in the literature are listed in Table 2. 

{\sf L6O:}\, 2MASXJ00140398-2310555 = NGC 45: the 2MASS source is a bright \HII\ region in the outer disc of this large
($D_{25}$=\am{7}{4}) spiral, whose E-W extent far exceeds the \am{3}{6} \nan\ HPBW. 
This explains why our integrated \HI\ line flux of 66 \Jykms, is much lower than the 
mean literature value of 245 \Jykms\ (LEDA), which was measured with larger telescope 
beams covering the entire galaxy.

{\sf L29O:}\, 2MASXIJ0041576-325810 = ESO 351-G2: our \HI\ detection 
($V_{HI}$=9560 \kms, $W_{50}$=193: \kms)  appears to have its high-velocity edge
reduced by negative RFI, and it may
be confused by a nearby galaxy. The target galaxy ($V_{opt}$=9618$\pm$49 \kms, $B_T$=15.27, 
$D_{25}$=\am{0}{9}), has an edge-on companion at \am{1}{1} separation, 
MCG-6-2-24 ($V_{opt}$=9525$\pm$60 \kms, $B_T$=16.55, $D_{25}$=\am{0}{6});
both are classified as Sc spirals. Both galaxies have a systemic velocity 
corresponding to the \HI\ velocity, within the errors, but the companion is
1.3 mag weaker.

{\sf L57O:}\, 2MASXJ01190447-0008190 = UGC 847: our \HI\ detection ($V_{HI}$=5238 \kms, $W_{50}$=271 \kms) 
may be slightly confused by a nearby galaxy. The edge-on target galaxy, 
($V_{opt}$=5218$\pm$60 \kms, $B_T$=16.29, $D_{25}$=\am{1}{4}, Scd) has a companion 
\am{3}{5} East of it, i.e. well outside the telescope's HPBW, PGC 212709 
($V_{opt}$=5173$\pm$22 \kms\ [Sloane Survey], $B_T$=16.27, $D_{25}$=\am{0}{6}).

{\sf L59O:}\, 2MASXJ01222287-3636297 = ESO 352-G58: our \HI\ profile may in principle be confused,
as at only \am{0}{5} from the centre of the target galaxy, 
($B_T$=15.82, $D_{25}$=\am{1}{2}, Sbc), lies the much smaller Irr galaxy PGC 634708 
($B_T$=17.75, $D_{25}$=\am{0}{35}). Neither has a known optical 
redshift, however. 

{\sf L70O:}\, 2MASXJ01430297-3411143 = IC 1722: though our \HI\ profile looks rather complex, 
with $V_{HI}$=3911 \kms\ 
and $W_{50}$=515 \kms, it does not seem likely it is confused with two nearby 
early-type galaxies. The target galaxy ($V_{opt}$=4082$\pm$35 \kms,
$B_T$=14.68, $D_{25}$=\am{1}{5}, SBbc), is highly inclined, as are nearby IC 1724 
($V_{opt}$=3836$\pm$51 \kms, $B_T$=14.53, $D_{25}$=\am{1}{2}, S0),
\am{3}{6} to the south, and ESO 353-G36 ($V_{opt}$=3774$\pm$137 \kms, $B_T$=15.11, 
$D_{25}$=\am{1}{0}, SO-a), \am{3}{4} to the east, outside the HPBW. Although the 
systemic velocities of the latter two objects are closer to the \HI\ velocity, they are 
both classified as lenticulars and therefore not expected to be gas-rich.

{\sf L86O:}\, 2MASXJ01590735-0308587 = MCG -1-6-13: \am{1}{1} east of the target galaxy
MCG -1-6-13 ($V_{opt}$=5342$\pm$42 \kms, $B_T$=14.71, $D_{25}$=\am{1}{0}, Irr), lies 
an Sa spiral of unknown velocity, KUG 0156-033 ($B_T$= 15.22, $D_{25}$=\am{0}{55}, Sa).

{\sf L117O:}\, 2MASXJ02284546-1030501 = NGC 948: our \HI\ profile of this object ($V_{HI}$=4758 \kms, 
$W_{50}$=168 \kms, $I_{HI}$=5.0 \Jykms) is probably confused with that of its 
larger neighbour NGC 945. At \am{2}{5} from the target galaxy ($V_{opt}$=4511$\pm$43 \kms, 
$B_T$=14.38, $D_{25}$=\am{1}{2}, SBc), lies the larger spiral NGC  945 
($V_{opt}$=4484$\pm$60 \kms, $B_T$=12.89, $D_{25}$=\am{2}{2}, SBc). At \nan\ 
Theureau et al. (1998) detected NGC 945 
($V_{HI}$=4482 \kms, $W_{50}$=179 \kms, $I_{HI}$=4.0 \Jykms), while at Arecibo 
Haynes et al. (1999) detected NGC 948 ($W_{50}$=181 \kms, $I_{HI}$=6.0 \Jykms). 
The E-W HPBW of the two telescopes is quite similar, \am{3}{6}, and serious confusion 
is inevitable between the line signal of both galaxies, whose centres are separated by \am{2}{5}.

{\sf (L125O):}\, 2MASXi J0235204+405522 = NGC 980: This object is classified as an S0 and therefore unlikely to 
be detectable in \HI, and our \HI\ spectrum ($V_{HI}$=5744 \kms, 
$W_{50}$=592 \kms, $I_{HI}$=4.0 \Jykms) is confused by nearby galaxies. 
Within the same telescope beam as the target object ($V_{opt}$=5796$\pm$42 \kms, 
$B_T$=14.27, $D_{25}$=\am{1}{7}), lies the Sa spiral NGC  982 ($V_{opt}$=5845$\pm$52 \kms, 
$B_T$=13.16, $D_{25}$=\am{1}{5}) at \am{3}{3}, the Sd spiral UGC 2068 ($B_T$=17.68, 
$D_{25}$=\am{0}{75}) at \am{2}{1} and the SBb spiral MCG +7-6-40 ($B_T$=17.96, 
$D_{25}$=\am{0}{35}) at \am{4}{2} distance; the latter two do not have a known redshift.
NGC 980 was detected at Arecibo ($V_{HI}$=5757 \kms, $W_{50}$=702 \kms,
$I_{HI}$=8.2 \Jykms; Haynes et al. 1988), NGC  982 also at Arecibo 
($V_{HI}$=5737 \kms, $W_{50}$=568 \kms, $I_{HI}$=7.0 \Jykms; Magri 1994) 
and UGC 2068 at \nan\ ($V_{HI}$=5740 \kms; Theureau et al. 1998). Seen the
similar optical redshifts of NGC  980 and 982 and their proximity all published \HI\ profiles of 
these objects must be confused.

{\sf L161O:}\, 2MASXIJ0306290-094332 = IC 1880: it is not clear to which galaxy our \HI\ detection 
($V_{HI}$=3935 \kms, $W_{50}$=144 \kms, $I_{HI}$=9.2 \Jykms)
is due. The target galaxy  is a lenticular at $V_{opt}$=10,224$\pm$42 \kms.
There is one other galaxy in the beam, of unknown redshift, 
PGC  135017 ($B_T$=16, $D_{25}$=\am{0}{7}, Sc) and another outside the HPBW,
NGC  1208 ($V_{opt}$=4518$\pm$53 \kms, $B_T$=13.31, $D_{25}$=\am{1}{9}),
which has a too high redshift and is classified as SO/a. Only a weak \HI\ line was detected 
towards NGC  1208 at \nan\ by Theureau et al. (1998) : $V_{HI}$=4356 \kms, $W_{50}$=259 \kms, 
$I_{HI}$=\am{0}{5} \Jykms.

{\sf L184O\, 2MASXJ03234832+4033485 = UGC 2708: it is not clear if the marginal \HI\ detection 
($V_{HI}$=8582: \kms, $W_{50}$=295: \kms, $I_{HI}$=6.5: \Jykms) is (mainly) due to residual 
RFI from the GPS L3 band centered on 8250 \kms\ (see Section 3.2.1), or to a galaxy.
It is $\sim$3200 \kms\ higher than the optical redshift of the target galaxy, 5390$\pm$51 \kms. 
Furthermore the galaxy has been classified as SB0 and is therefore unlikely to be detectable. 
There are no likely candidates for confusion in the vicinity. We therefore consider this a tentative
detection only.}

{\sf L203O:}\, 2MASXJ03404154-2239041 = ESO 482-G31: our narrow \HI\ detection ($V_{HI}$=1683 \kms, 
$W_{50}$=60 \kms, $I_{HI}$=0.5 \Jykms) may be due to a nearby galaxy.
The target galaxy ($V_{opt}$ 1629$\pm$43 \kms, $B_T$=15.27,
 $D_{25}$=\am{0}{9}) is classified as an S0. At the edge of the beam lies the much 
larger NGC  1415 ($V_{opt}$=1549$\pm$47 \kms, $B_T$=12.77, $D_{25}$=\am{3}{3}). 
Though classified as S0/a, it has quite a strong \HI\ line as detected, respectively, 
at \nan\ (Balkowski 1979) and Effelsberg (Huchtmeier 1982): mean values 
$V_{HI}$=1585 \kms, $W_{50}$=322 \kms\ and $I_{HI}$=8.3 \Jykms.

{\sf  L237O:}\, 2MASXJ04294085-2724313 = NGC 1592:  This Irregular object appears to be a merger or a very close, 
interacting pair of galaxies.

{\sf L249O:}\, 2MASXIJ0443334-274006 = ESO 421-G13: our \HI\ spectrum shows two peaks, at 
$V_{HI}$=5353 and 10,457 \kms, respectively. The target galaxy ($B_T$=15.3, 
$D_{25}$=\am{1}{2}, Sbc) does not have a known optical redshift; 
there is no other candidate for confusion within the beam in the search area.
Seen its size and magnitude, an association with the 5353 \kms\ profile seems
the most likely.

{\sf L255O:}\, 2MASXJ04511837-0557379 = MCG -01-13-021: One nearby galaxy, 2MASXi J0451123-060013, 
lies within the beam, while two others lie on the edge of the beam: MCG -01-13-020
($B_T$=15.29, $D_{25}$=\am{1}{0}) and 2MASXi J0451085-060321. 
Like the target galaxy ($B_T$=15.23, $D_{25}$=\am{1}{5}, Sc), none of these 
have a known redshift.

{\sf L256O:}\, 2MASXJ04551547-1209275 = MCG -2-13-23: It is not clear to which of the two galaxies in the telescope
beam galaxy our detection ($V_{HI}$=4862  \kms, $W_{50}$=231 \kms, \IHI=4.4 \Jykms) is due,
as the target galaxy has an optical redshift of  17,881$\pm$60  
\kms\ (Huchra et al. 1993), which is far outside our velocity search range, while the other galaxy 
in the beam, MCG -2-13-22, is an S0 with $V_{opt}$= 8133 $\pm$60  \kms.
The observed line signal may be spurious and due imperfectly filtered out RFI.

{\sf L272O:}\, 2MASXJ05153881-2242320 = ESO 486-G40  : It seems unlikely that our detection 
is confused by the one magnitude weaker AM 0513-224, of unknown redshift, at the edge
of the beam. Our \HI\ redshift of the target galaxy, 5796 \kms, corresponds to its optical 
value of 5779$\pm$42 \kms.

{\sf L297O:}\, 2MASXJ06131885+5306445 = UGC 3421:  It is not clear if our \HI\ detection of the target galaxy 
($D_{25}$=\am{0}{9}, Sd) may be confused by a galaxy at the edge of the beam 
(at \am{3}{3} EW separation), UGC 3424  ($B_T$=16.24, $D_{25}$=\am{1}{2}, Irr), as 
neither has a published redshift.

{\sf L302O:}\, 2MASXJ06243891-2235497 = ESO 489-G50:  The target galaxy ($V_{opt}$=2892 \kms, $B_T$=15.79, 
$D_{25}$=\am{1}{4}), is classified as S0-a and therefore unlikely to be gas-rich. 
Our \HI\ detection ($V_{HI}$=2713 \kms, $W_{50}$=316 \kms, \IHI=9.8 \Jykms) appears 
due to a large galaxy at the edge of the beam, \am{14}{4} south of the target galaxy, 
NGC  2223 ($V_{opt}$=2690$\pm$61 \kms, $B_T$=12.51, $D_{25}$=\am{2}{9}, SBc), 
whose mean \HI\ profile parameters 
are $V_{HI}$=2721 \kms, $W_{50}$=300 \kms\ and \IHI=26 \Jykms\ (Bottinelli et al. 1982; 
Fisher \& Tully 1981; Staveley-Smith \& Davies 1987).

{\sf L336O:}\, 2MASXIJ0819021+211122 = UGC 4329: the redshift of our \HI\ detection 
($V_{HI}$=4079 \kms, $W_{50}$=222 \kms, $I_{HI}$=6.8 \Jykms) is in agreement with the 
$V_{opt}$=4144$\pm$73 \kms, and the spectrum is unlikely to be confused by the 4 galaxies 
that lie at the edge of the beam (CGCG 119-043, CGCG 119-053,  NGC  2556 and NGC  2557) 
as they  all have redshifts in the range of 4460-4860 \kms. Its line parameters agree with the 
published mean values of
$V_{HI}$=4097 \kms, $W_{50}$=223 \kms\ and $I_{HI}$=8.7 \Jykms\ 
(Bicay \& Giovanelli 1986; Bothun et al. 1985; Lewis 1983; Lewis et al 1985; 
Rosenberg \& Schneider 2000; Schommer et al. 1981; Tifft \& Cocke 1988).

{\sf L345O:}\, 2MASXJ08325749-2254031 = ESO 495-G17: our \HI\ detection  ($V_{HI}$=1416 \kms, $W_{50}$=205 \kms, 
$I_{HI}$=3.2 \Jykms) 
may be due to gas in the outer disc of the nearby large spiral NGC 2613. Another \nan\ 
spectrum (Chamaraux et al. 
1999) of our target galaxy ($B_T$=14.76, $D_{25}$=\am{1}{4}, Sb), of 
unknown redshift, shows  $W_{50}$=178 \kms\ and $I_{HI}$=3.9 \Jykms.
Although the centre of NGC 2613 ($V_{opt}$=1599$\pm$88 \kms, $B_T$=11.15,  
$D_{25}$=\am{7}{0}) lies \am{7}{3} SE of that of the target galaxy, its diameter is large 
enough to cause confusion.
\HI\ observations of  NGC  2613 (Bottinelli et al. 1982; Fisher \& Tully 1981;
Reif et al. 1982; Staveley-Smith \& Davies 1987) show mean parameters
of $V_{HI}$=1678 \kms, $W_{50}$=599 \kms\ and $I_{HI}$=83 \Jykms.

{\sf L358O:}\, 2MASXJ08565629-2031383 = ESO 563-G34: our narrow \HI\ profile (($V_{HI}$=2616 \kms, $W_{50}$=97 \kms, 
$I_{HI}$=4.4 \Jykms) 
of  the target galaxy ($V_{opt}$=2558$\pm$60 \kms, $B_T$=14.32, $D_{25}$=\am{1}{2}, Sab) does 
not appear to be confused by a galaxy on the edge of the beam, at $\sim$3$'$ E-W 
separation, ESO 563-G36 ($V_{opt}$=2656$\pm$50 \kms, $B_T$=13.88, $D_{25}$=\am{1}{4}, Sab). 

{\sf  L360O:}\, 2MASXJ09001445+3543527 = NGC 2719:  Confused \HI\ spectrum. The target galaxy 
($V_{opt}$=3172$\pm$46 \kms, $B_T$=13.65, $D_{25}$=\am{1}{2}, Irr),
forms a close pair (\am{0}{5} separation) with
NGC  2719A ($V_{opt}$=3069$\pm$43 \kms, $B_T$=14.50, $D_{25}$=\am{0}{55}, Irr).
Surprisingly, our \HI\ detection ($V_{HI}$=3043 \kms, $W_{50}$=215 \kms, 
$I_{HI}$=9.3 \Jykms) has a  
114 \kms\ lower central velocity and a two times smaller integrated line flux than 
the Arecio profile of Bicay 
\& Giovanelli (1986), with $V_{HI}$=3157 \kms, $W_{50}$=231 \kms\ and $I_{HI}$=17.9Jy \kms, 
which also includes both galaxies. 

{\sf L371O:}\, 2MASXIJ0917317+415932 = NGC 2799: our \HI\ profile is confused, as there are two other sizeable galaxies of 
similar redshift in the beam of the target galaxy
($V_{opt}$=1860$\pm$68 \kms, $B_T$=13.98, $D_{25}$=\am{1}{8}, SBd): 
NGC  2798 ($V_{opt}$=1735$\pm$67 \kms, $B_T$=14.92, $D_{25}$=\am{2}{7}, SBa) and
UGC  4904 ($V_{opt}$=1644$\pm$60 \kms, $B_T$=15.21, $D_{25}$=\am{1}{0}, SBbc).
The E-W separation between NGC  2799/98 is \am{1}{6}, i.e. only about half the \nan\ HPBW. 
Our \HI\ profile has $V_{HI}$=1616 \kms, $W_{50}$=159 \kms\ and $I_{HI}$=9.1 \Jykms, while 
the \nan\ profile of NGC 2799 from Bottinelli et al. (1982) has $V_{HI}$=1755 \kms, 
$W_{50}$=340 \kms\ and $I_{HI}$=11.1 \Jykms\  and a  \nan\ profile of NGC  2798 
(Bottinelli et al. 1980) 
has $V_{HI}$=1726 \kms, $W_{50}$=313 \kms\ and $I_{HI}$=11.5  \Jykms.
Our profile has a  $\sim$125 \kms\ lower central velocity and a two times smaller 
$W_{50}$. Profiles that include the three above-mentioned galaxies were published by
Huchtmeier (1982) and Peterson \& Shostak (1974).
% {\bf These show xxxxx.}

{\sf (F26O)}  = MCG -1-25-33: our \HI\ profile is probably confused, since there is a 
pair of galaxies, 
of unknown redshift, of similar size in the beam of the target galaxy 
($B_T$=14.48, $D_{25}$=\am{1}{3}, SBcd):
UGC A 173 ($V_{opt}$=1875$\pm$97 \kms, $B_T$=14.63, $D_{25}$=\am{1}{3}, SBd)
and UGC A 174 ($B_T$=14.85, $D_{25}$=\am{1}{3}, SBm).
Our \HI\ profile ($V_{HI}$=1883 \kms, $W_{50}$=118 \kms, $I_{HI}$=11.0 \Jykms)  
agrees with the Richter \& Huchtmeier (1987) Effelsberg profile of UGC A 173 
($V_{HI}$=1867 \kms, $W_{50}$=189 \kms, $I_{HI}$=12.7 Jy\kms), which includes 
all three objects. 

{\sf L392O:}\, 2MASXJ09450315-0429475 = FGC 936: there are two galaxies, both of unknown redshift, with only 
\am{0}{6} separation at the centre of the beam, FGC 936 ($B_T$=16.6, $D_{25}$=\am{0}{9}, Sc) 
and 2MASXi J0945014-042919 (=PGC 1057506), with $B_T$=16.45 and $D_{25}$=\am{0}{5}). 
While in NED the position 
of our 2MASS source coincides exactly with that of FGC  936, 
in LEDA it coincides exactly with that of the other galaxy. As the LEDA 
positions all have an accuracy of 10$''$, significantly better than that from the FGC 
(Karachentsev et al. 1999), we adopted the latter association.       

{\sf L403O:}\, 2MASXJ10144367-2048367 = ESO 567-G28: our \HI\ profile ($V_{HI}$=3705 \kms, $W_{50}$=208 \kms, 
$I_{HI}$=2.9 \Jykms) may in principle be confused. At about the HPBW level of our beam, 
pointed at the target galaxy ($B_T$=15.43, $D_{25}$=\am{1}{2}, Sbc) lie two galaxies, 
also of unknown redshift: 
ESO 567-G28 ($B_T$=15.87, $D_{25}$=\am{0}{6}, Sbc) and 2MASXi J1014531-205752 
(=PGC 834527), with $B_T$=16.19 and $D_{25}$=\am{0}{8}.

{\sf L447O:}\, 2MASXJ11174025+3803095 = UGC 6307: our profile has a 58 \kms\ lower central velocity than the Green Bank 
91m profile of Schneider et al. (1992), although the $W_{50}$ line widths are the same (206 \kms). 
The \nan\ integrated line flux is 1.4 times higher. 
No candidates for confusion within the telescope beams are visible on the DSS.

{\sf L455O:}\, 2MASXIJ1128100+165531 = NGC 3691: our profile ($V_{HI}$=987 \kms)
has an on average 100 \kms\ lower central velocity
than the 4 published \HI\ spectra, three of which were measured at Arecibo and one with the
Green Bank 91m telescope (see Table 2). The line widths and integrated line fluxes are
in agreement, however. The optical systemic velocity of 997$\pm$38 \kms\ (Blackman 1980)
is in good agreement with our centre velocity.
The \nan\ profile has a peculiar shape and may in principle have been affected by RFI.
No candidates for confusion within the telescope beams are visible on the DSS.

{\sf L459O:}\, 2MASXJ11375027+4752539 = NGC 3769A: the \HI\ profile of the target galaxy is confused with that of 
NGC  3769, the much larger spiral with which it forms a close pair. Our \HI\ profile 
($V_{HI}$=734 \kms, $W_{50}$=230 \kms, $I_{HI}$=35.0 \Jykms) is in agreement with 
the mean of the literature values of $V_{HI}$=732 \kms, $W_{50}$=232 \kms\ and 
$I_{HI}$=50.6 \Jykms\  (Magri 1994; Rhee \& van Albada 1996; Verheijen \& Sancisi 2001). 
The latter refers to a mapping of the pair with the WSRT.

{\sf L481O:}\, 2MASXJ11562491-1210003 = MCG -02-31-002: our \HI\ spectrum shows two peaks, at 
$V_{HI}$=5429 and 11,891 \kms, respectively. 
The latter detection appears to have its high-velocity edge clipped by negaative RFI.
The target galaxy ($B_T$=15.4, $D_{25}$=\am{1}{2}) 
does not have a known optical redshift and there is no other galaxy in the search area likely to cause 
confusion in the Nan\c{c}ay beam. 

{\sf L497O:}\, 2MASXIJ1212205+291240 = NGC 4173:  Our \HI\ spectrum shows two detections, at 
$V_{HI}$=1054 and 3795 \kms, respectively. The former is due to the target galaxy, 
at  $V_{opt}$=1117$\pm$32 \kms, while the latter is due to 
three other galaxies in the beam (NGC 4169, 4174 and 4175), which have optical 
redhifts in the range of 3780-4000 \kms. The redshift of our profile of NGC  4173 
($V_{HI}$=1054 \kms, $W_{50}$=163 \kms, $I_{HI}$=27.6 \Jykms) is 53 \kms\ higher 
than the mean LEDA values ($V_{HI}$=1107 \kms, $W_{50}$=116 \kms, $I_{HI}$=13.4 \Jykms) 
and its integrated line flux is twice as high.

{\sf L502P:}\, 2MASXIJ1216418-251505 = PGC 781011: 
two peaks are seen in our \HI\ spectrum, at 451 and 5708 \kms, 
respectively. On the DSS no obvious candidate could be found that might correspond to the 
former \HI\ signal ($V_{HI}$=451 \kms, $W_{50}$=65 \kms, $I_{HI}$=1.1 \Jykms). We assume 
that the latter peak ($V_{HI}$=5708 \kms, $W_{50}$=62 \kms, $I_{HI}$=1.6 \Jykms) 
corresponds to the target galaxy ($B_T$=16.1, $D_{25}$=\am{0}{5}).

{\sf L548O:}\, 2MASXJ12503459-0931108 = MCG -01-33-022: our \HI\ profile of the target galaxy 
($V_{HI}$=4689 \kms, $W_{50}$=274 \kms, $I_{HI}$=4.0 \Jykms), with an optical redshift of 
4666$\pm$80 \kms, is very probably confused with that of a nearby pair of galaxies, 
NGC  4716/7. NGC  4716 is an S0 with $V_{opt}$=4573$\pm$119 \kms\ and NGC  4717 is a spiral 
with $V_{opt}$=4515$\pm$96 kms; none of them has previously been reported as 
detected in HI.

{\sf L569O:}\, 2MASXJ13165624-1635347 = MCG -03-34-040: the target galaxy is seen 
superposed on the outermost regions 
of a much larger Sbc spiral, NGC  5054, which will dominate the \HI\ profile. Our \HI\ line parameters 
($V_{HI}$=1682 \kms, $W_{50}$=321 \kms, $I_{HI}$=14.8 \Jykms) are quite different the mean published 
values of NGC 5054 $V_{HI}$=1741 \kms, $W_{50}$=315 \kms, $I_{HI}$=26.8 \Jykms\ (Bottinelli et al. 1982; 
Fisher \& Tully  1981; Richter \& Huchtmeier 1987; Shostak 1978).

{\sf L582O:}\, 2MASXJ13324730+4151564 = ESO 579-G5: the target galaxy is seen 
superposed on the outer regions of a much larger 
Sc spiral, NGC  5214, which will dominate the \HI\ profile. Our \HI\ line parameters 
($V_{HI}$=8174 \kms, $W_{50}$=303 \kms, $I_{HI}$=7.6 \Jykms) are similar to the published values of 
NGC  5124 $V_{HI}$=8180 \kms, $W_{50}$=306 \kms, $I_{HI}$=4.9 \Jykms\ (Haynes \& Giovanelli 1991). 
In the same beam as the target galaxy ($B_T$=15.8,  $D_{25}$=\am{1}{0}, Sbc), lies the somewhat 
fainter Sab ESO 579-G6 ($B_T$=16.2, 
$D_{25}$=\am{0}{8}). Since neither has a known optical redshift, it 
cannot be determined if their proximity could cause confusion in our \HI\ 
spectrum.

{\sf (F40O)}  = MCG -2-36-13: the radial velocity of the detected line profile, 4882 \kms, is completely 
different from the optical value of 12,500 \kms\ (da Costa et al. 1998), which lies far outside
our velocity search range. No candidate was found for confusion within 
the beam. The observed line signal may be spurious and due imperfectly filtered to RFI.

{\sf L643O:}\, 2MASXIJ1447561-141658 = MCG-02-38-015: the target object corresponds to an outer spiral arm 
of the spiral MCG-02-38-015. Our 
\HI\ profile parameters ($V_{HI}$=1927 \kms, $W_{50}$=221 \kms, $I_{HI}$=22.3 Jy 
\kms) show a 66 \kms\ lower mean velocity than 
the average of the literature values, $V_{HI}$=2046 \kms, $W_{50}$=199 \kms\ and 
$I_{HI}$=23.4 \Jykms\  (Theureau et al. 1998; Thonnard et al. 1978).        

{\sf L647O:}\, 2MASXJ14563333-2430079 = ESO 513-G009: our \HI\ spectrum shows two peaks, at 
$V_{HI}$=6894 and 8600 \kms, respectively. The optical redshift of the target galaxy,  
6815$\pm$44 \kms, corresponds to the lower velocity peak. No galaxy could be 
identified that may have caused the other peak.

{\sf L651O:}\, 2MASXJ15064639+1251009 = CGCG 077-007: the \HI\ profile of the target galaxy,  
($V_{opt}$=6582$\pm$105\kms, $B_T$=16.1, $D_{25}$=\am{0}{7}) will be confused with that 
of NGC  5851 ($V_{opt}$=6515$\pm$64 \kms, $B_T$=14.2, 
 $D_{25}$=\am{1}{0}), at \am{1}{8} E of the 2MASS object, i.e. around the Nan\c{c}ay HPBW. 
Our \HI\ profile shows $V_{HI}$=6576 \kms, $W_{50}$=43 \kms\ and $I_{HI}$=5.5 \Jykms, while 
the Nan\c{c}ay profile of NGC 5851 
(Bottinelli et al. 1999) shows $V_{HI}$=6508 \kms, $W_{50}$=281 \kms\ and $I_{HI}$=2.3 \Jykms.

{\sf L654P:}\, 2MASXJ15205260-2859139 = PGC 160462: our \HI\ spectrum shows two peaks, at $V_{HI}$=3934 and 6022 \kms, 
respectively. The target galaxy has no known optical redshift and no other galaxy was found 
in the search area. It is not clear with which of the two \HI\ lines it is associated.

{\sf L705O:}\, 2MASXIJ1719223+490202 = VV 010b: our \HI\ spectrum ($V_{HI}$=7478 \kms, $W_{50}$=97 \kms, 
$I_{HI}$=2.6 \Jykms) will be confused by that of the interacting pair Arp 102 (PGC  60067/70); 
the centre of  PGC  60070 lies only \am{0}{5} from that of the much smaller Sd target galaxy, 
PGC 60073 ($B_T$=16.2, $D_{25}$=\am{0}{7})
PGC 60070 is an Sd with $V_{opt}$=7177$\pm$35 \kms, $B_T$=15.5 and  
$D_{25}$=\am{1}{1}, and PGC 60067 an elliptical with $V_{opt}$=7304$\pm$99 \kms, 
$B_T$=15.5 and $D_{25}$=\am{0}{95}.

{\sf L712O:}\, 2MASXJ18030213+2922257 = CGCG 171-049: the \HI\ spectrum ($V_{HI}$=7048 \kms, $W_{50}$=473 \kms, 
$I_{HI}$=3.3 \Jykms) of the target galaxy ($V_{opt}$=7048$\pm$46 \kms, $B_T$=15.7, 
$D_{25}$=\am{0}{8}) may be confused by another galaxy in the telescope beam, 
CGCG 171-048 ($V_{opt}$=6983$\pm$44 \kms, $B_T$=15.4, 
$D_{25}$=\am{0}{75}).

{\sf L745O:}\, 2MASXJ20170275-1206516 = MCG -02-51-007: the \HI\ spectrum ($V_{HI}$=5656 \kms, $W_{50}$=355 \kms, 
$I_{HI}$=5.8 \Jykms) of the target galaxy ($B_T$=15.6, log(D$_{25}$=\am{0}{8}, Sb), without 
known optical redshift, may be confused by another galaxy in the telescope beam, MCG -02-51-008 
($V_{opt}$=5578$\pm$60 \kms, $B_T$=14.3, $D_{25}$=\am{0}{6}, Sc).

{\sf L752O:}\,  = FGC 2286: our profile has a 43 \kms\ lower central velocity than the \nan\ 
profile of Matthews \& van Driel (2000), although the line widths and integrated
line fluxes are comparable (see Table 2).

{\sf L755O:}\, 2MASXJ20481195-0352186 = MCG -01-53-009: the \HI\ spectrum ($V_{HI}$=6042 \kms, $W_{50}$=128 \kms, 
$I_{HI}$=1.2 \Jykms) of the target 
galaxy ($V_{opt}$=5945$\pm$60 \kms, B=13, $D_{25}$=\am{1}{1}), may 
in principle be confused by an object of similar size but lower surface brightness at only \am{0}{9} 
separation, MCG -01-53-010 (B=15, $D_{25}$=\am{1}{0}), without known redshift.

{\sf L758O:}\, 2MASXJ20550682-3105304 = ESO 463-IG34: the target galaxy is a close interacting pair of galaxies.

{\sf L766O:}\, 2MASXJ21455799-3448541 = ESO 403-G23: our \HI\ spectrum shows two peaks, at $V_{HI}$=5140 and 12,924 \kms, 
respectively. No other galaxy was found in the search area besides the target galaxy ($B_T$=15.4, 
$D_{25}$=\am{1}{2}, Sc), without known optical redshift. Seen the galaxy's optical parameters, it is 
most likely with the \HI\ emission at 5140 \kms.

{\sf L777P:}\, 2MASXIJ2216009-095612 = PGC 984498: the target galaxy ($B_T$=16.5, $D_{25}$=\am{0}{5}), shows two nuclei 
on the DSS and 2MASS images, as well as a tidal tail on the DSS. There is another galaxy of similar 
magnitude in the beam, 2MASXi J2216005-095324 ($B_T$=16.5, $D_{25}$=\am{0}{9}), also without 
optical redshift. 

{\sf L783O:}\, 2MASXJ22230550-2857173 = NGC 7259: the \HI\ spectrum ($V_{HI}$=1657 \kms, $W_{50}$=152 \kms, 
$I_{HI}$=11.6 \Jykms) of the target galaxy ($V_{opt}$=1718$\pm$42 \kms, $B_T$=13.6, 
$D_{25}$=\am{1}{2}), will be confused by an edge-on galaxy at \am{2}{8} distance, 
ESO 467-G51 ($V_{opt}$=1775$\pm$32 \kms, $B_T$=14.6, $D_{25}$=\am{2}{8}), of which 
Theureau et al. (1998) obtained an \HI\ profile at Nan\c{c}ay with 
$V_{HI}$=1808 \kms, $W_{50}$=146 \kms, $I_{HI}$=18.6 \Jykms.
      
{\sf L794O:}\, 2MASXIJ2244347-225930 = ESO 534-G21: the \HI\ profile of the target galaxy will not be 
confused by its close apparent companion, ESO-LV 5340211 at only \am{0}{3} distance, as their 
optical redshifts are 3159 and 18,116 \kms, respectively.

\section{Discussion}  % 5
A comparison of the global parameters of the \HI\ profiles measured for our survey
at \nan\ with those obtained from the literature is given in Table 2. Indicated with a
`*' after the source name in column 1 are the cases in which profile(s) are likely 
to be confused with those of other galaxies in the telescope beam(s), 
see Section 4.5 for further details. 
We also compared the line parameters of the 18 calibrator galaxies observed throughout
the survey.

Plotted in Figures 4, 5 and 6 are, respectively, the differences between central line velocities
from our survey and from the literature, survey and literature $W_{50}$
line widths and $I_{HI}$ integrated line fluxes for the 2MASS survey galaxies with non-confused 
spectra as well as for the calibration galaxies.

The agreement between the central \HI\ velocities is generally good (Fig. 4).
The 4 survey galaxies for which the literature values are (on average) more than 
40 \kms\ different from our values are 
L447O (UGC 6307), L455O (NGC 3691), L643O (MCG 2-38-15) and L752O (FGC 2286),
see Section 4.5 for comments. In general no clear cause can be identified for
these discrepancies, although RFI in our data may be involved in some cases.

The agreement between the $W_{50}$ line widths is generally good (Fig. 5), with two
exceptions. These two discrepancies, of about -80 \kms, for L497O (NGC 4173) and calibration
galaxy NGC 3321 (PGC 31653) are due to one anomalously narrow published line width per
galaxy, while all  other literature values are in agreement with ours.

A comparison between the integrated line fluxes $I_{HI}$ (Fig. 6) shows a larger
scatter for the survey objects than the two other plots, which appears reasonable, however, seen 
the dispersion found in other such
plots (e.g., van Driel et al. 2000). The three cases in which a literature value
exceed our measurement by more than a factor 2 are L6O (NGC 45), which is
considerably larger than the \nan\ beam, L445O (IC 2627), which may
well have been resolved by the \nan\ beam and L487O (UGC 7000), due to the 
anomalously high Arecibo value of Impey et al. (1996).

The fluxes of the calibrator galaxies are in good agreement with the literature values.
Comparing to measurements made with other telescopes, which are available for 7 galaxies, 
we find that our $I_{HI}$ values are on average 0.95$\pm$0.25 times these values. 
Comparing to the \nan\ measurements of 18 galaxies by Theureau et al. (1998), which were 
made with the previous receiver system (see Section 3), we find a scaling factor of 1.10$\pm$0.22. 
The standard deviations in these scaling factors are only slightly larger than the $\sim$$\pm$15\% 
accuracy of the internal calibration of the \nan\ data (e.g., Matthews et al. 2001).

There are 4 survey galaxies (L180O = FGC 410, L467O = UGC 6679, L630O = ESO 447-5 and
L673O = MCG 3-41-136) that did not meet our criteria for clear or marginal detections, 
which were reported as detected in the literature (see Table 2). All four were observed
by us in the low-velocity search mode only, i.e. in the -500 to 10,500 \kms\ range,
which means that the systemic velocity of L630O (10,600 \kms) lies just outside our search range.
The other three were detected in the \nan\ survey of Matthews \& van Driel (2000),
which has a significantly lower noise level than the present survey.

\section{Conclusions}  % 6
Of the 334 galaxies observed in total, 171 (51\%) were clearly detected,
76 of which did not have a previously known radial velocity.
The detection statistics as function of type (previously catalogued (PGC/other catalogues) or not)
and size/selection algorithm (Large/Faint) are listed in table 1. 
For the Large objects ($r_{K_{20}}\geq 20''$) and the Faint objects (selected
with the LCSB source processor), the overall detection rates are quite
similar: 51\% and 56\%, respectively.

\acknowledgements{ 
This publication makes use of data products from the Two Micron 
All Sky Survey, which is a joint project of the University of Massachusetts 
and the Infrared Processing and Analysis Center, funded by the National 
Aeronautics and Space Administration and the National Science Foundation.
The \nan\ Radio Observatory, which is the Unit\'e Scientifique de
\nan\ of the Observatoire de Paris, is associated with the French 
Centre National de Recherche Scientifique (CNRS) as USR B704,
and acknowledges the financial support of the R\'egion Centre as well as of 
the European Union. 
This research also has made use of the Lyon-Meudon Extragalactic 
Database (LEDA), recently incorporated in HyperLeda, 
the NASA/IPAC Extragalactic Database (NED)   
which is operated by the Jet Propulsion Laboratory, California Institute   
of Technology, under contract with the National Aeronautics and Space      
Administration and the Aladin database, operated at CDS, Strasbourg, France.  
We acknowledge financial support from CNRS/NSF collaboration grant No.
10637.
}

\onecolumn
\newpage

% -------------------------- Table 1 -------------------------------
\begin{table}
\centering
\bigskip
{\footnotesize
\begin{tabular}{llllllll}
\multicolumn{8}{l}{{\bf Table 1.} \HI\ detection statistics }\\
\smallskip \\
\hline
\vspace{-2 mm} \\
 &  \multicolumn{2}{c} {Previously} & \multicolumn{2}{c} {Previously} & 
\multicolumn{2}{c} {Previously} & \multicolumn{1}{c} {All} \\ 
 &  \multicolumn{2}{c} {Catalogued} & \multicolumn{2}{c} {Catalogued} & 
\multicolumn{2}{c} {Uncatalogued} & \\
 &  \multicolumn{2}{c} {(Other)} & \multicolumn{2}{c} {(PGC)}& \multicolumn{2}{c} 
{(New)} & \\  
 &  \multicolumn{1}{c} {Large} & \multicolumn{1}{c} {Faint} 
 &  \multicolumn{1}{c} {Large} & \multicolumn{1}{c} {Faint}   
 &  \multicolumn{1}{c} {Large} &  \multicolumn{1}{c} {Faint} & \\   
\vspace{-2 mm} \\
\hline
\vspace{-2 mm} \\
Total observed & 229      & 14     &   41 & 0    & 39     & 11    & 334 \\
\vspace{-2 mm} \\
 Detected      &     140 61\% &       10  71\% & 12 29\% & 0 0\% &  \,\,5  13\% & 4  36\% & 171 51\%  \\
 Marginal      &  \,\,\,23 10\% &  \,\,\,0 \,\,0\% & \,\,\,0  \,\,0\% & 0 0\% &  \,\,0  \,\,\,0\% & 0 \,\,\,0\% &  \,\,24\,
\,\,7\%  \\
 Undetected   &  \,\,\,66 29\% & \,\,\,4  29\% & 29 71\% & 0 0\% & 34  87\% & 7  64\% &  139 42\%  \\
\vspace{-2 mm} \\
\hline
\vspace{-2 mm} \\
\multicolumn{8}{l}{{\bf Note:} Large: galaxies with $r_{K_{20}}\geq 20''$, Faint: galaxies selected using the} \\
\multicolumn{8}{l}{ LCSB source processor.} \\
\end{tabular}
}
\normalsize
\end{table}

% \twocolumn
\centering

% -------------------------- Table 2 -------------------------------
\begin{table}
\bigskip
{\scriptsize
\begin{tabular}{lllrrrll}
\multicolumn{8}{l}{\footnotesize {\bf Table 2.} Comparison with published \HI\ data}\\
\smallskip \\
\hline
\vspace{-2 mm} \\     
No.   & PGC & Ident. & $V_{HI}$ & $W_{50}$ & $I_{HI}$ & Tel. & Ref. \\
      &     &        &  km/s      & km/s &   {\tiny Jy km/s} &      &      \\
\vspace{-2 mm} \\
\hline 
\vspace{-1 mm} \\
\multicolumn{8}{l}{Clearly detected galaxies} \\
\end{tabular}
}
\end{table}

% -------------------------- Table 2 -------------------------------
\begin{table}
\bigskip
{\scriptsize
\begin{tabular}{lllrrrll}
\multicolumn{8}{l}{\footnotesize {\bf Table 2.} Comparison with published \HI\ data  -- {\it continued}}\\
\smallskip \\
\hline
\vspace{-2 mm} \\     
No.   & PGC & Ident. & $V_{HI}$ & $W_{50}$ & $I_{HI}$ & Tel. & Ref. \\
      &     &        &  km/s      & km/s &   {\tiny Jy km/s} &      &      \\
\vspace{-2 mm} \\
\hline 
\vspace{-1 mm} \\
\multicolumn{8}{l}{Clearly detected galaxies -- continued} \\
\end{tabular}
}
\end{table}

% --------------- Table 3a.1 --------------------------------------------
%
\begin{table}
\centering
\bigskip
{\scriptsize
\begin{tabular}{llrl}
\multicolumn{4}{l}{\footnotesize {\bf Table 3a.1.} Detected objects } \\
\multicolumn{4}{l}{\footnotesize   \,\,\,\,\,\,\,\,\,\,\,\,\,\,\,\,\,\,\,\,\,\,\,\,\,\,\,\, 
                                  -- identifications and coordinates}\\
\smallskip \\
\hline
\vspace{-2 mm} \\
Number & 2MASS & PGC & others  \\ 
& hhmmss.s+ddmmss &  &  \\  
 & (J2000.0) &  &  \\ 
\vspace{-2 mm} \\
\end{tabular}
}
\normalsize
\end{table} 

% --------------- Table 3a.2 --------------------------------------------
%
\begin{table}
\centering
\bigskip
{\scriptsize
\begin{tabular}{llrl}
\multicolumn{4}{l}{\footnotesize {\bf Table 3a.2.} Detected objects}\\
\multicolumn{4}{l}{\footnotesize \,\,\,\,\,\,\,\,\,\,\,\,\,\,\,\,\,\,\,\,\,\,  
                                      -- identifications and coordinates - {\it continued} }\\
\smallskip \\
\hline
\vspace{-2 mm} \\
Number & 2MASS & PGC & others  \\ 
& hhmmss.s+ddmmss &  &  \\  
 & (J2000.0) &  &  \\ 
\vspace{-2 mm} \\
\end{tabular}
}
\normalsize
\end{table}

% --------------- Table 3a.3 --------------------------------------------
%
\begin{table}
\centering
\bigskip
{\scriptsize
\begin{tabular}{llrl}
\multicolumn{4}{l}{\footnotesize {\bf Table 3a.3.} Detected objects}\\
\multicolumn{4}{l}{\footnotesize \,\,\,\,\,\,\,\,\,\,\,\,\,\,\,\,\,\,\,\,\,\,  
                                      -- identifications and coordinates - {\it continued} }\\
\smallskip \\
\hline
\vspace{-2 mm} \\
Number & 2MASS & PGC & others  \\ 
& hhmmss.s+ddmmss &  &  \\  
 & (J2000.0) &  &  \\ 
\vspace{-2 mm} \\
\end{tabular}
}
\normalsize
\end{table} 

 % Tab 3a.1 - 3a.3  dets, idents
  
% --------------- Table 3b.1 --------------------------------------------
%
\begin{sidewaystable}
\centering
\bigskip
{\scriptsize
\begin{tabular}{lrrrrrrrrlrrrrrrrrrrrrr}
\multicolumn{23}{l}{\footnotesize {\bf Table 3b.1.} Detected objects -- basic data}\\
\smallskip \\
\hline
\vspace{-2 mm} \\
No & $K_{20}$ & $J-K$ & $\mu_{K5}$ & $b/a$ & $r_{K_{20}}$ & $B_{T_{c}}$ & $D_{25}$ & $\mu_{B_{25}}$ & $T$ & rms & 
 $I_{HI}$ & $V_{HI}$  & $\sigma_{V_{HI}}$ & $V_{opt}$ & $W_{50}$ & $W_{20}$ & $D$ & \MHI\  & \LK\ & \LB\ &  \MHILKfr\ &  \MHILBfr\  \\
 & &  &  &  &  &  &  &  &  &  &  &  &  &  &  &  &  & [log]  & [log]  & [log]  &  &  \\
    & mag & mag  & $\frac{mag}{''^2}$  &  & $''$ & mag & $'$ & $\frac{mag}{''^2}$ &   & mJy & 
 \JYKMS\ & \KMS\ & \KMS\ & \KMS\ & \KMS\ & \KMS\ & Mpc & \Msun\ & \LsunK\ & \LsunB\  & \multicolumn{2}{c}{solar units} \\ 
\vspace{-2 mm} \\
\hline
\vspace{-2 mm} \\
\end{tabular}
}
\normalsize
\end{sidewaystable}

% --------------- Table 3b.2 --------------------------------------------
\begin{sidewaystable}
\centering
\bigskip
{\scriptsize
\begin{tabular}{lrrrrrrrrlrrrrrrrrrrrrr}
\multicolumn{23}{l}{\footnotesize {\bf Table 3b.2.} Detected objects -- basic data - {\it continued} }\\
\smallskip \\
\hline
\vspace{-2 mm} \\
No & $K_{20}$ & $J-K$ & $\mu_{K5}$ & $b/a$ & $r_{K_{20}}$ & $B_{T_{c}}$ & $D_{25}$ & $\mu_{B_{25}}$ & $T$ & rms & 
 $I_{HI}$ & $V_{HI}$  & $\sigma_{V_{HI}}$ & $V_{opt}$ & $W_{50}$ & $W_{20}$ & $D$ & \MHI\  & \LK\ & \LB\ &  \MHILKfr\ &  \MHILBfr\  \\
 & &  &  &  &  &  &  &  &  &  &  &  &  &  &  &  &  & [log]  & [log]  & [log]  &  &  \\
    & mag & mag  & $\frac{mag}{''^2}$  &  & $''$ & mag & $'$ & $\frac{mag}{''^2}$ &   & mJy & 
 \JYKMS\ & \KMS\ & \KMS\ & \KMS\ & \KMS\ & \KMS\ & Mpc & \Msun\ & \LsunK\ & \LsunB\  & \multicolumn{2}{c}{solar units} \\ 
\vspace{-2 mm} \\
\hline
\vspace{-2 mm} \\
\end{tabular}
}
\normalsize
\end{sidewaystable}

% --------------- Table 3b.3 --------------------------------------------
\begin{sidewaystable}
\centering
\bigskip
{\scriptsize
\begin{tabular}{lrrrrrrrrlrrrrrrrrrrrrr}
\multicolumn{23}{l}{\footnotesize {\bf Table 3b.3.} Detected objects -- basic data - {\it continued} }\\
\smallskip \\
\hline
\vspace{-2 mm} \\
No & $K_{20}$ & $J-K$ & $\mu_{K5}$ & $b/a$ & $r_{K_{20}}$ & $B_{T_{c}}$ & $D_{25}$ & $\mu_{B_{25}}$ & $T$ & rms & 
 $I_{HI}$ & $V_{HI}$  & $\sigma_{V_{HI}}$ & $V_{opt}$ & $W_{50}$ & $W_{20}$ & $D$ & \MHI\  & \LK\ & \LB\ &  \MHILKfr\ &  \MHILBfr\  \\
 & &  &  &  &  &  &  &  &  &  &  &  &  &  &  &  &  & [log]  & [log]  & [log]  &  &  \\
    & mag & mag  & $\frac{mag}{''^2}$  &  & $''$ & mag & $'$ & $\frac{mag}{''^2}$ &   & mJy & 
 \JYKMS\ & \KMS\ & \KMS\ & \KMS\ & \KMS\ & \KMS\ & Mpc & \Msun\ & \LsunK\ & \LsunB\  & \multicolumn{2}{c}{solar units} \\ 
\vspace{-2 mm} \\
\hline
\vspace{-2 mm} \\
\end{tabular}
}
\normalsize
\end{sidewaystable}

% --------------- Table 3  b.4 --------------------------------------------
\begin{sidewaystable}
\centering
\bigskip
{\scriptsize
\begin{tabular}{lrrrrrrrrlrrrrrrrrrrrrr}
\multicolumn{23}{l}{\footnotesize {\bf Table 3b.4.} Detected objects -- basic data - {\it continued} }\\
\smallskip \\
\hline
\vspace{-2 mm} \\
No & $K_{20}$ & $J-K$ & $\mu_{K5}$ & $b/a$ & $r_{K_{20}}$ & $B_{T_{c}}$ & $D_{25}$ & $\mu_{B_{25}}$ & $T$ & rms & 
 $I_{HI}$ & $V_{HI}$  & $\sigma_{V_{HI}}$ & $V_{opt}$ & $W_{50}$ & $W_{20}$ & $D$ & \MHI\  & \LK\ & \LB\ &  \MHILKfr\ &  \MHILBfr\  \\
 & &  &  &  &  &  &  &  &  &  &  &  &  &  &  &  &  & [log]  & [log]  & [log]  &  &  \\
    & mag & mag  & $\frac{mag}{''^2}$  &  & $''$ & mag & $'$ & $\frac{mag}{''^2}$ &   & mJy & 
 \JYKMS\ & \KMS\ & \KMS\ & \KMS\ & \KMS\ & \KMS\ & Mpc & \Msun\ & \LsunK\ & \LsunB\  & \multicolumn{2}{c}{solar units} \\ 
\vspace{-2 mm} \\
\hline
\vspace{-2 mm} \\
\hline															     			
\end{tabular}														     
}															     
\normalsize														     
\end{sidewaystable}

% --------------- Table 3b.5 --------------------------------------------
\begin{sidewaystable}
\centering
\bigskip
{\scriptsize
\begin{tabular}{lrrrrrrrrlrrrrrrrrrrrrr}
\multicolumn{23}{l}{\footnotesize {\bf Table 3b.5.} Detected objects -- basic data - {\it continued} }\\
\smallskip \\
\hline
\vspace{-2 mm} \\
No & $K_{20}$ & $J-K$ & $\mu_{K5}$ & $b/a$ & $r_{K_{20}}$ & $B_{T_{c}}$ & $D_{25}$ & $\mu_{B_{25}}$ & $T$ & rms & 
 $I_{HI}$ & $V_{HI}$  & $\sigma_{V_{HI}}$ & $V_{opt}$ & $W_{50}$ & $W_{20}$ & $D$ & \MHI\  & \LK\ & \LB\ &  \MHILKfr\ &  \MHILBfr\  \\
 & &  &  &  &  &  &  &  &  &  &  &  &  &  &  &  &  & [log]  & [log]  & [log]  &  &  \\
    & mag & mag  & $\frac{mag}{''^2}$  &  & $''$ & mag & $'$ & $\frac{mag}{''^2}$ &   & mJy & 
 \JYKMS\ & \KMS\ & \KMS\ & \KMS\ & \KMS\ & \KMS\ & Mpc & \Msun\ & \LsunK\ & \LsunB\  &  \multicolumn{2}{c}{solar units} \\ 
\vspace{-2 mm} \\
\hline
\vspace{-2 mm} \\
\end{tabular}
}
\normalsize
\end{sidewaystable}

 % Tab 3b.1 - 3b.5  dets, basic data
\begin{table}
\centering
\bigskip
{\scriptsize
\begin{tabular}{llrl}
\multicolumn{4}{l}{\footnotesize {\bf Table 4a.} Marginal detections } \\
\multicolumn{4}{l}{\footnotesize   \,\,\,\,\,\,\,\,\,\,\,\,\,\,\,\,\,\,\,\,\,\,\,\,\,\,\,\, 
                                  -- identifications and coordinates}\\
\smallskip \\
\hline
\vspace{-2 mm} \\
Number & 2MASS & PGC & others  \\ 
& hhmmss.s+ddmmss &  &  \\  
 & (J2000.0) &  &  \\
\vspace{-2 mm} \\
\end{tabular}
}
\normalsize
\end{table} 

\begin{sidewaystable}
\centering
\bigskip
{\scriptsize
\begin{tabular}{lrrrrrrrrlrrrrrrrrrrr}
\multicolumn{21}{l}{\footnotesize {\bf Table 4b.} Marginal detections -- basic data}\\
\smallskip \\
\hline
\vspace{-2 mm} \\
No & $K_{20}$ & $J-K$ & $\mu_{K5}$ & $b/a$ & $r_{K_{20}}$ & $B_{T_{c}}$ & $D_{25}$ & $\mu_{B_{25}}$ & $T$ & rms & 
 $I_{HI}$ & $V_{HI}$  & $V_{opt}$ & $W_{50}$ & $D$ & \MHI\  & \LK\ & \LB\ &  \MHILKfr\ &  \MHILBfr\  \\
 &  &  &  &  &  &  &  &  &  &  &  &  &  &  &  & [log]  & [log]  & [log]  &  &  \\
    & mag & mag & $\frac{mag}{''^2}$  &  & $''$ & mag & $'$ & $\frac{mag}{''^2}$ &   & mJy & 
 \JYKMS\  & \KMS\ & \KMS\ & \KMS\ & Mpc & \Msun\ & \LsunK\ & \LsunB\  &   & \\ 
\vspace{-2 mm} \\
\hline
\vspace{-2 mm} \\
\end{tabular}
}
\normalsize
\end{sidewaystable}

 % Tab 4a-b           marginal dets
\begin{table}                                                                                                                                                                                           
\centering                                                                                                                                                                                              
\bigskip                                                                                                                                                                                                
{\scriptsize                                                                                                                                                                                            
\begin{tabular}{llrl}                                                                                                                                                                                  
\multicolumn{4}{l}{\footnotesize {\bf Table 5a.1.} Undetected objects}\\                                                                                                                       
\multicolumn{4}{l}{\footnotesize  \,\,\,\,\,\,\,\,\,\,\,\,\,\,\,\,\,      
                                  -- identifications and coordinates}\\                                                                                                                                
\smallskip \\                                                                                                                                                                                           
\hline                                                                                                                                                                                                  
\vspace{-2 mm} \\                                                                                                                                                                                       
Number & 2MASS & PGC & others  \\ 
& hhmmss.s+ddmmss &  &  \\  
 & (J2000.0) &  &  \\ 
\vspace{-2 mm} \\
\end{tabular}                                                                                                                                                                                           
}                                                                                                                                                                                                       
\normalsize                                                                                                                                                                                             
\end{table}                                                                                                                                                                                             
\begin{table}                                                                                                                                                                                           
\centering                                                                                                                                                                                              
\bigskip                                                                                                                                                                                                
{\scriptsize                                                                                                                                                                                            
\begin{tabular}{llrl}                                                                                                                                                                                  
\multicolumn{4}{l}{\footnotesize {\bf Table 5a.2.} Undetected objects}\\                                                                                                                       
\multicolumn{4}{l}{\footnotesize  \,\,\,\,\,\,\,\,\,\,\,\,\,\,\,\,\,      
                                  -- identifications and coordinates - {\it continued}}\\                                                                                                                                
\smallskip \\                                                                                                                                                                                           
\hline                                                                                                                                                                                                  
\vspace{-2 mm} \\                                                                                                                                                                                       
Number & 2MASS & PGC & others  \\ 
& hhmmss.s+ddmmss &  &  \\  
 & (J2000.0) &  &  \\ 
\vspace{-2 mm} \\
\end{tabular}                                                                                                                                                                                           
}                                                                                                                                                                                                       
\normalsize                                                                                                                                                                                             
\end{table}

\begin{sidewaystable}                                                                                                                                                                                   
\centering                                                                                                                                                                                              
\bigskip                                                                                                                                                                                                
{\scriptsize                                                                                                                                                                                            
\begin{tabular}{lrrrrrrrrlccccl}                                                                                                                                                                       
\multicolumn{15}{l}{\footnotesize {\bf Table 5b.1.} Undetected objects -- basic data }\\                                                                                         
\smallskip \\                                                                                                                                                                                           
\hline                                                                                                                                                                                                  
\vspace{-2 mm} \\                                                                                                                                                                                       
No & $K_{20}$ & $J-K$ & $\mu_{K5}$ & $b/a$ & $r_{K_{20}}$ & $B_{T_{c}}$ & $D_{25}$ & $\mu_{B_{25}}$ & $T$ & rms & $I_{HI}$ &    
  \MHILKfr\  &   \MHILBfr\  & search \\          
    & mag & mag & $\frac{mag}{''^2}$  &  & $''$ & mag & $'$ & $\frac{mag}{''^2}$ &  & mJy & \JYKMS\   & \multicolumn{2}{c}{solar units} & mode \\         
\vspace{-2 mm} \\                                                                                                                                                                                       
\hline                                                                                                                                                                                                  
\vspace{-2 mm} \\
\hline                                                                                                                                                                                                  
\end{tabular}                                                                                                                                                                                           
}                                                                                                                                                                                                       
\normalsize                                                                                                                                                                                             
\end{sidewaystable}

% ---------------------- Table 5b.2 -------------------------------------   
%                                                                                                                                                                                                         
\begin{sidewaystable}                                                                                                                                                                                   
\centering                                                                                                                                                                                              
\bigskip                                                                                                                                                                                                
{\scriptsize                                                                                                                                                                                            
\begin{tabular}{lrrrrrrrrlccccl}                                                                                                                                                                       
\multicolumn{15}{l}{\footnotesize {\bf Table 5b.2.} Undetected objects 
                                   -- basic data - {\it continued}}\\                                                                                         
\smallskip \\                                                                                                                                                                                           
\hline                                                                                                                                                                                                  
\vspace{-2 mm} \\                                                                                                                                                                                       
No & $K_{20}$ & $J-K$ & $\mu_{K5}$ & $b/a$ & $r_{K_{20}}$ & $B_{T_{c}}$ & $D_{25}$ & $\mu_{B_{25}}$ & $T$ & rms & $I_{HI}$ & 
   \MHILKfr\  &   \MHILBfr\  & search \\          
    & mag & mag & $\frac{mag}{''^2}$  &  & $''$ & mag & $'$ & $\frac{mag}{''^2}$ &  & mJy &  \JYKMS\   & \multicolumn{2}{c}{solar units} & mode \\         
\vspace{-2 mm} \\                                                                                                                                                                                       
\hline                                                                                                                                                                                                  
\vspace{-2 mm} \\            
L316O &   11.98 &    1.04 &   18.19 &   0.63 & 20.4 & 14.86 &   0.79 &   23.53 &    Sd    &     4.8 &   3.6 &   0.25 &   0.48&  lo  \\
\hline                                                                                                                                                                                                  
\end{tabular}                                                                                                                                                                                           
}                                                                                                                                                                                                       
\normalsize                                                                                                                                                                                             
\end{sidewaystable}

% ---------------------- Table 5b.3 -------------------------------------   
%                                                                                                                                                                                                         
\begin{sidewaystable}                                                                                                                                                                                   
\centering                                                                                                                                                                                              
\bigskip                                                                                                                                                                                                
{\scriptsize                                                                                                                                                                                            
\begin{tabular}{lrrrrrrrrlccccl}                                                                                                                                                                       
\multicolumn{15}{l}{\footnotesize {\bf Table 5b.3.} Undetected objects 
                                   -- basic data - {\it continued}}\\                                                                                         
\smallskip \\                                                                                                                                                                                           
\hline                                                                                                                                                                                                  
\vspace{-2 mm} \\                                                                                                                                                                                       
No & $K_{20}$ & $J-K$ & $\mu_{K5}$ & $b/a$ & $r_{K_{20}}$ & $B_{T_{c}}$ & $D_{25}$ & $\mu_{B_{25}}$ & $T$ & rms & $I_{HI}$ & 
   \MHILKfr\  &   \MHILBfr\  & search \\          
    & mag & mag & $\frac{mag}{''^2}$  &  & $''$ & mag & $'$ & $\frac{mag}{''^2}$ &  & mJy &  \JYKMS\   & \multicolumn{2}{c}{solar units} & mode \\         
\vspace{-2 mm} \\                                                                                                                                                                                       
\hline                                                                                                                                                                                                  
\vspace{-2 mm} \\            
L691O &   12.39 &    1.02 &   18.17 &   0.26 & 21.0 & 15.68 &   0.73 &    23.79 &    Sc    &  5.0 &   3.7 &   0.37 &   1.05&  lo  \\
\end{tabular}                                                                                                                                                                                           
}                                                                                                                                                                                                       
\normalsize                                                                                                                                                                                             
\end{sidewaystable}                                                                                                                                                                                     
     
 % Tab 5a.1 - 5b.3  non-dets

% \begin{figure*} % --- Fig.  -----------------
% \centering
% \includegraphics{}
% \caption [] {Example of the application of the off-line radio frequency interference (RFI) mitigation 
% algorithm (see Section 3): the upper \HI\ spectrum was reduced without applying it, the lower with.
% {\bf -- to be done.}}
% \end{figure*}
% -----------------------------------------------------------------

% ----- FIGURES 2 and 3 ------------------------------------------------------
%
\begin{figure*} % --- 2a -----------------
\centering
\includegraphics{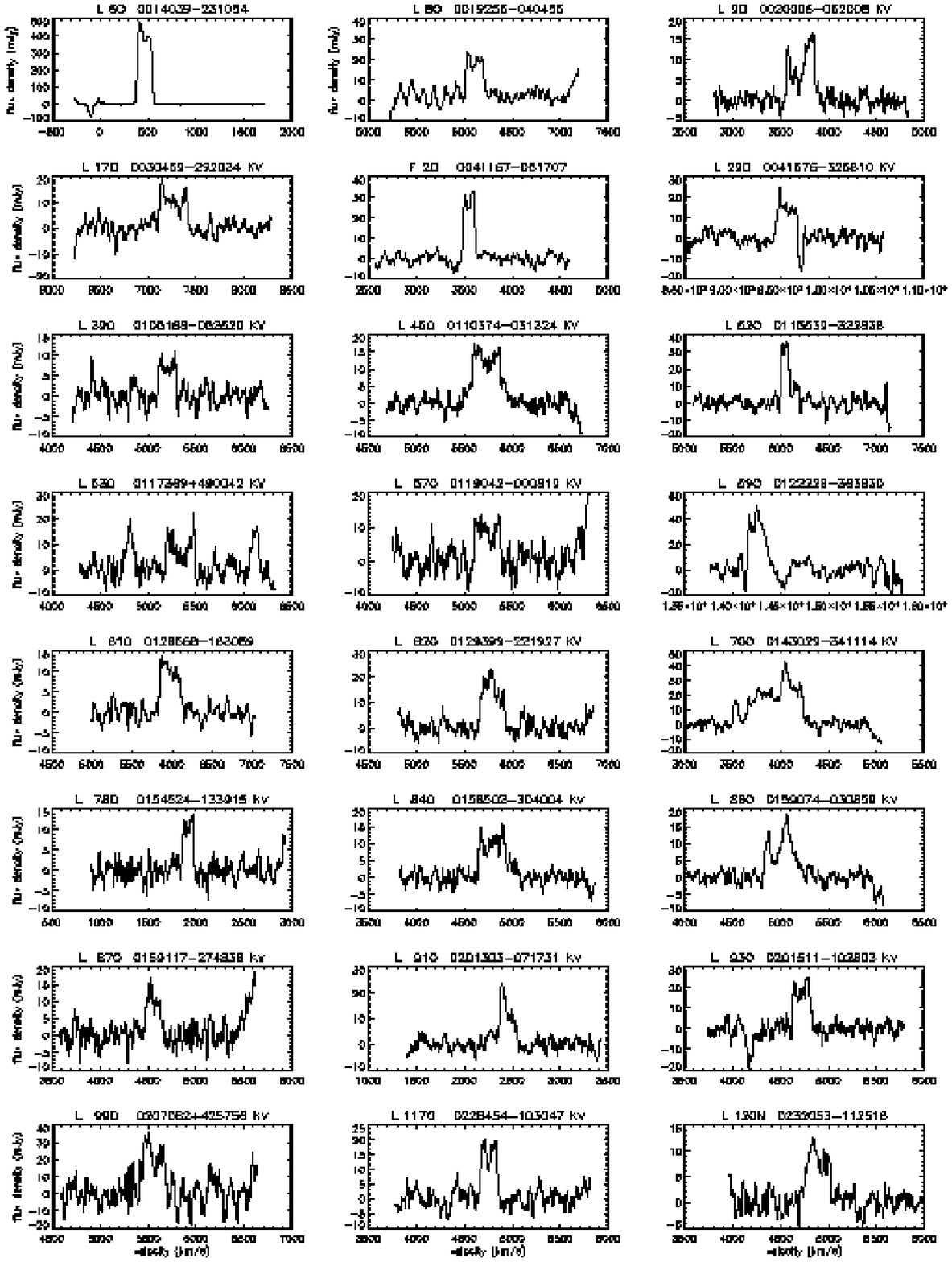}
\caption [] {{\bf a.}\, Nan\c{c}ay 21-cm \HI\ line spectra of detected objects (see Table 3). 
Velocity resolution is 15.8 \kms\ (velocity search mode) and 17.1 \kms\ (known velocity mode), 
radial heliocentric velocities are according to the optical convention. Galaxies detected in the 
`known velocity' mode are indicated by the designation 'KV' following their coordinates in the
header of their spectrum.
}
\end{figure*}

\begin{figure*} % --- 2b -----------------
\centering
\addtocounter{figure}{-1}
\includegraphics{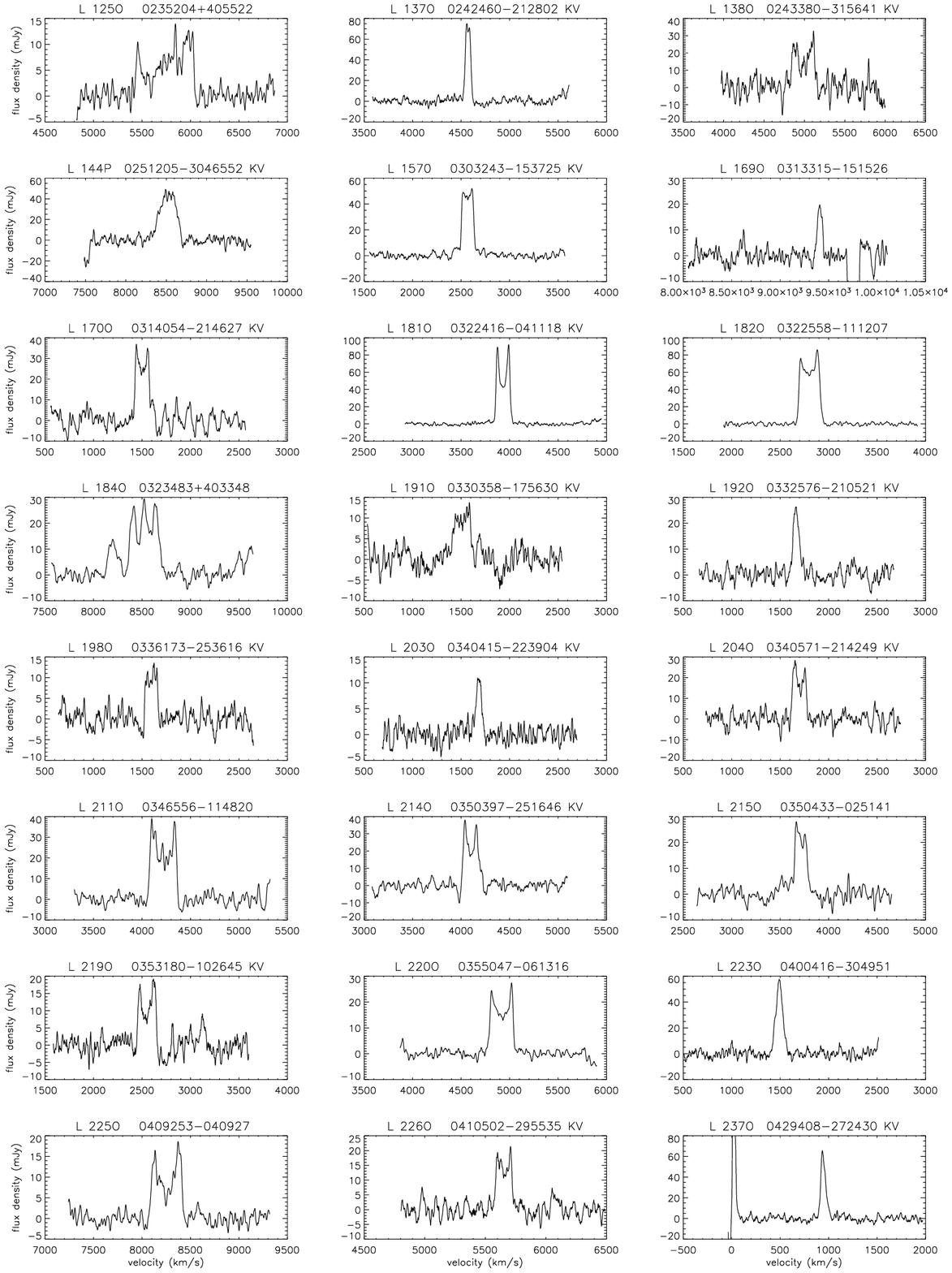}
\caption [] {{\bf b.}\, Nan\c{c}ay 21-cm \HI\ line spectra
of detected objects {\it - continued}.
}
\end{figure*}

\begin{figure*} % --- 2c -----------------
\centering
\addtocounter{figure}{-1}
\includegraphics{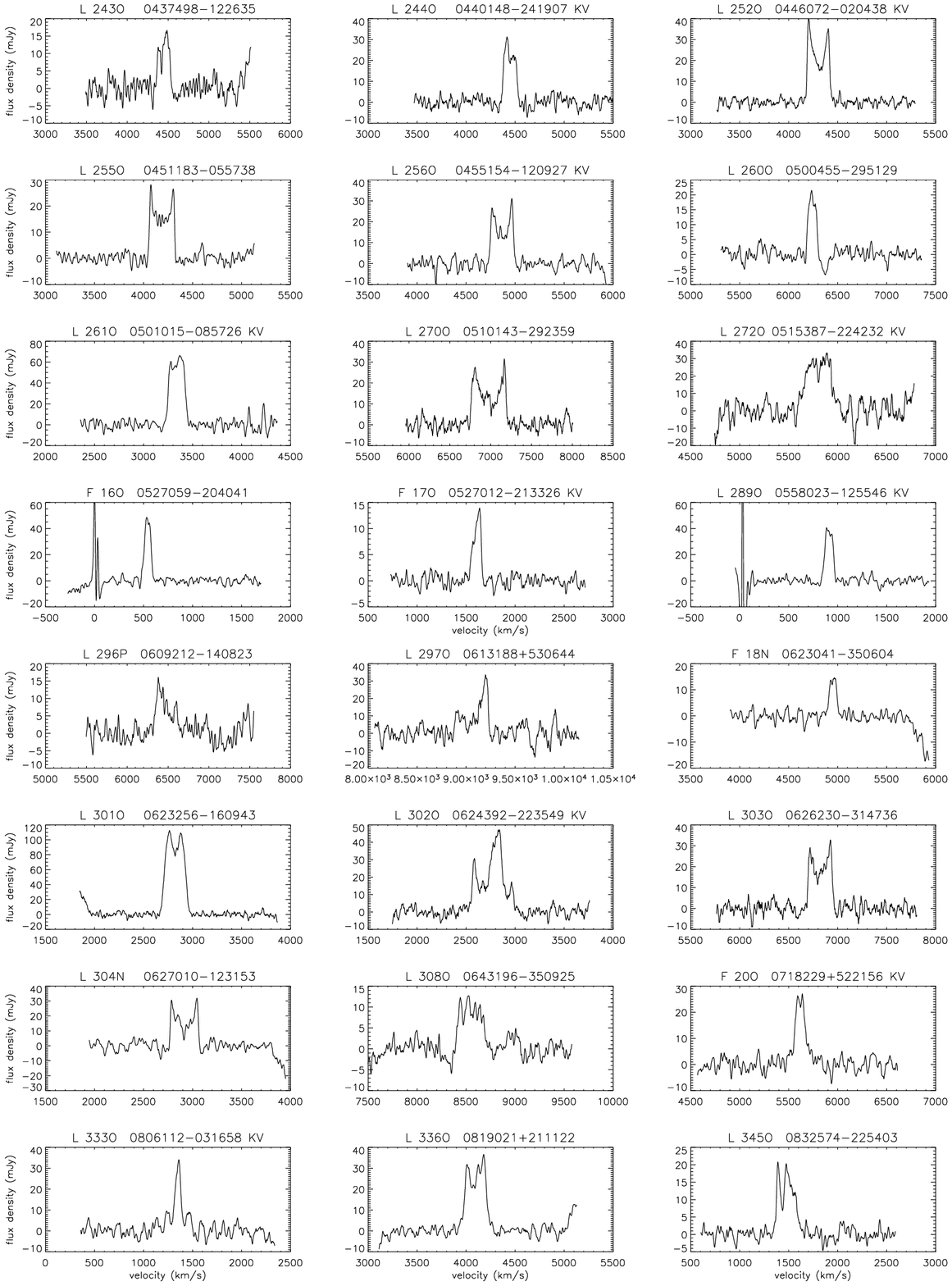}
\caption [] {{\bf c.}\, Nan\c{c}ay 21-cm \HI\ line spectra
of detected objects {\it - continued}. 
}
\end{figure*}

\begin{figure*} % --- 2d -----------------
\centering
\addtocounter{figure}{-1}
\includegraphics{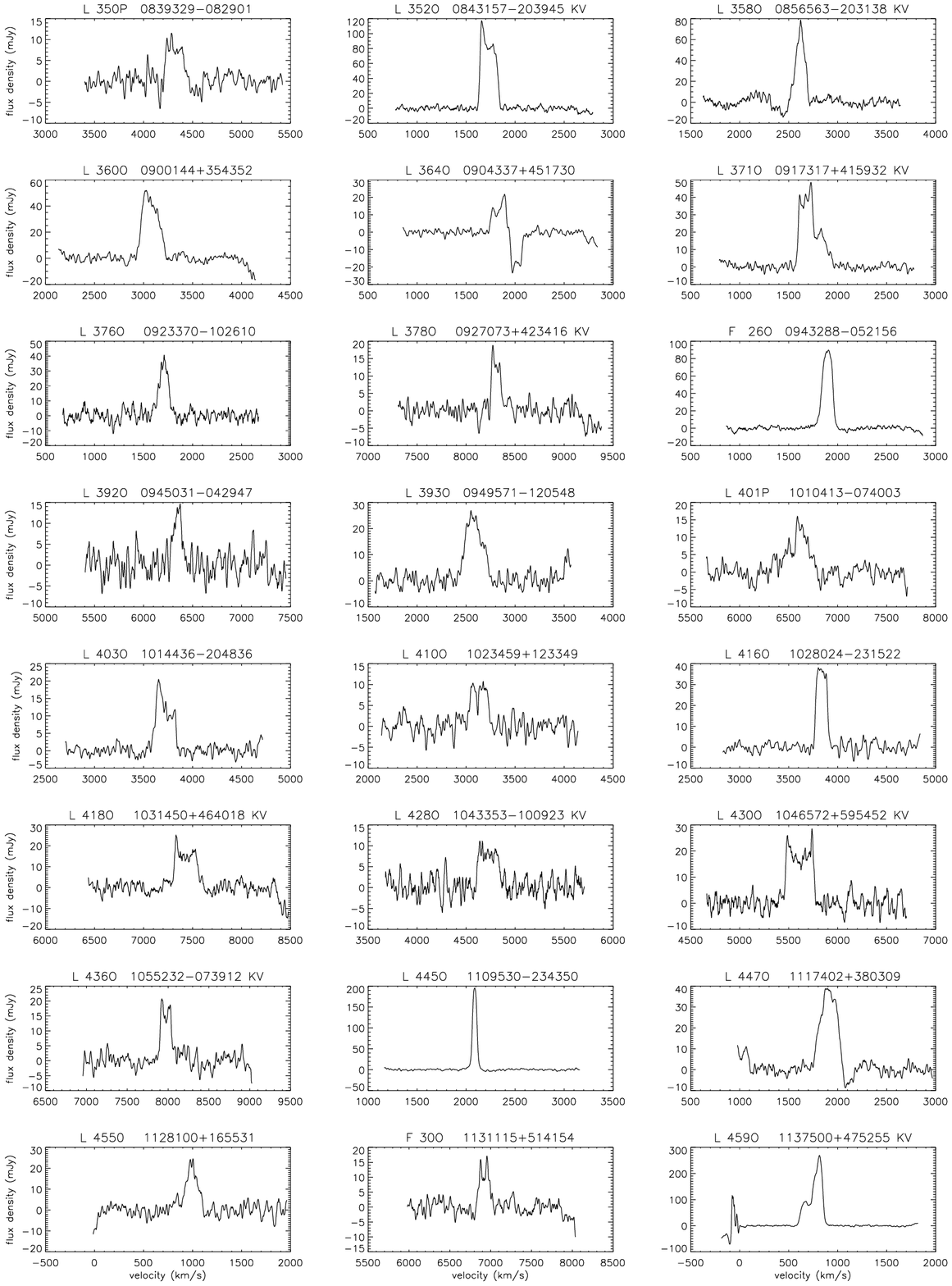}
\caption [] {{\bf d.}\, Nan\c{c}ay 21-cm \HI\ line spectra
of detected objects {\it - continued}. 
}
\end{figure*}

\begin{figure*} % --- 2e -----------------
\centering
\addtocounter{figure}{-1}
\includegraphics{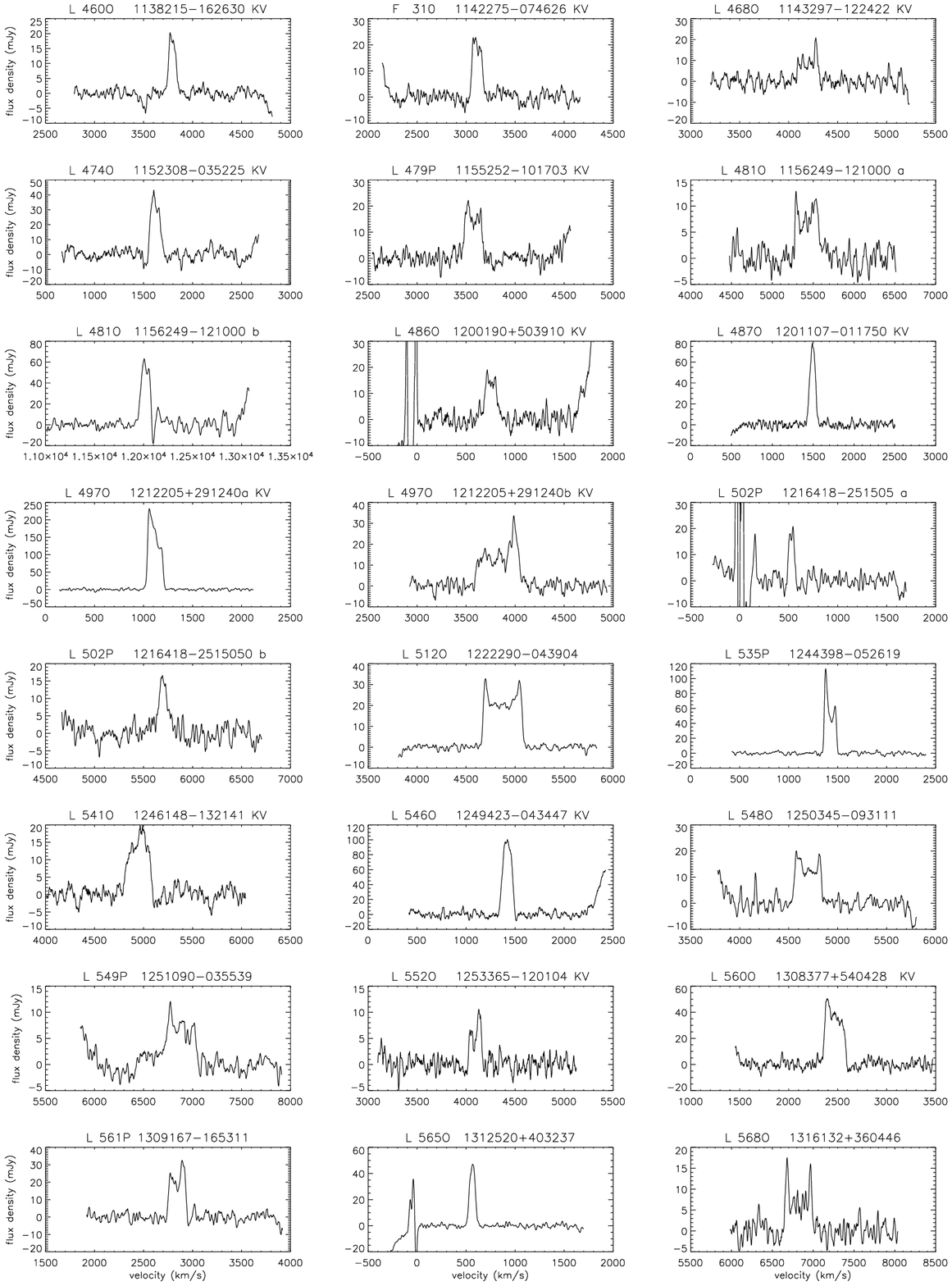}
\caption [] {{\bf e.}\, Nan\c{c}ay 21-cm \HI\ line spectra
of detected objects {\it - continued}. 
}
\end{figure*}

\begin{figure*} % --- 2f -----------------
\centering
\addtocounter{figure}{-1}
\includegraphics{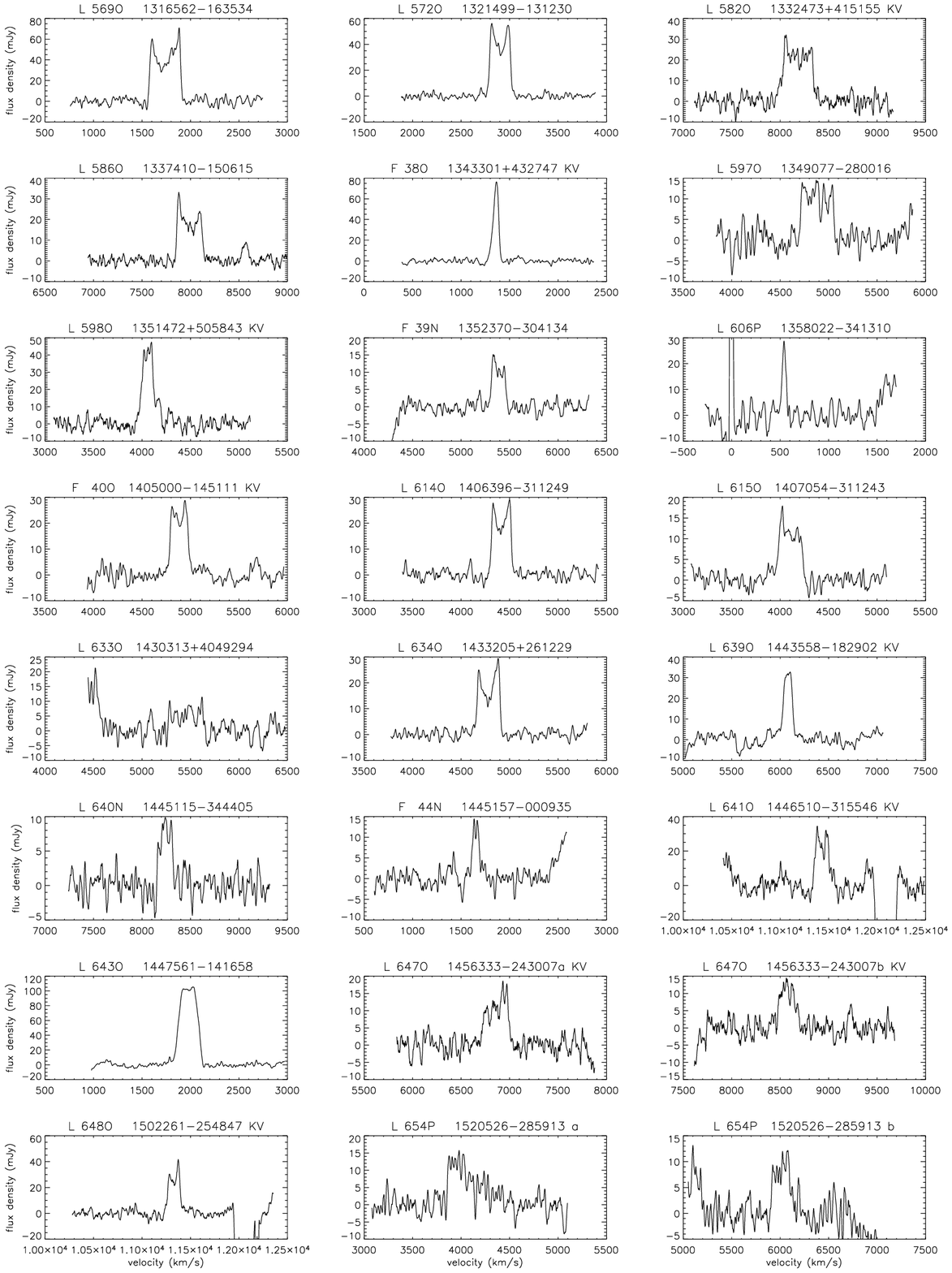}
\caption [] {{\bf f.}\, Nan\c{c}ay 21-cm \HI\ line spectra
of detected objects {\it - continued}. 
}
\end{figure*}

\begin{figure*} % --- 2g -----------------
\centering
\addtocounter{figure}{-1}
\includegraphics{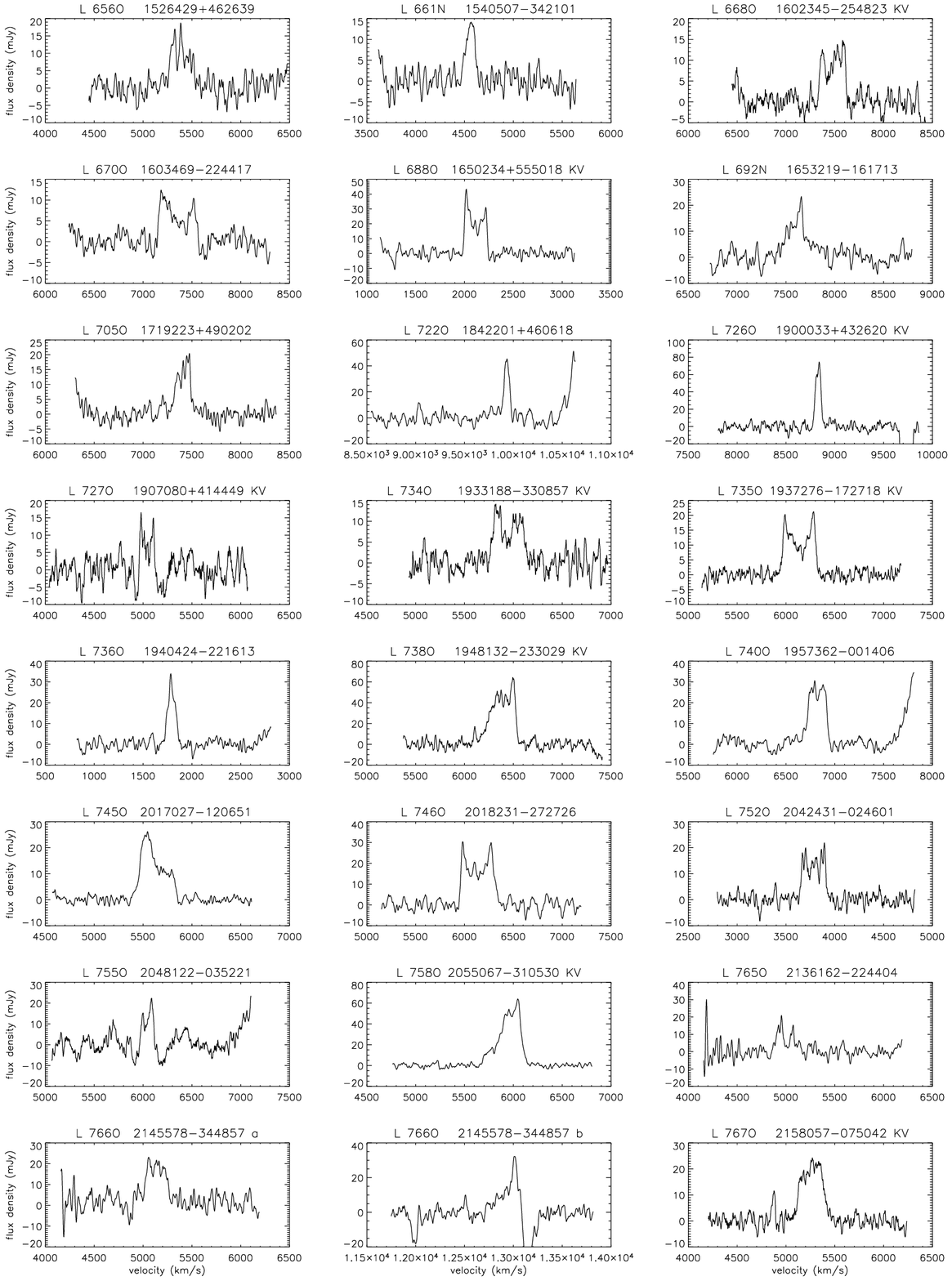}
\caption [] {{\bf g.}\, Nan\c{c}ay 21-cm \HI\ line spectra
of detected objects {\it - continued}. 
}
\end{figure*}
\begin{figure*} % --- 2h -----------------
\centering
\addtocounter{figure}{-1}
\includegraphics{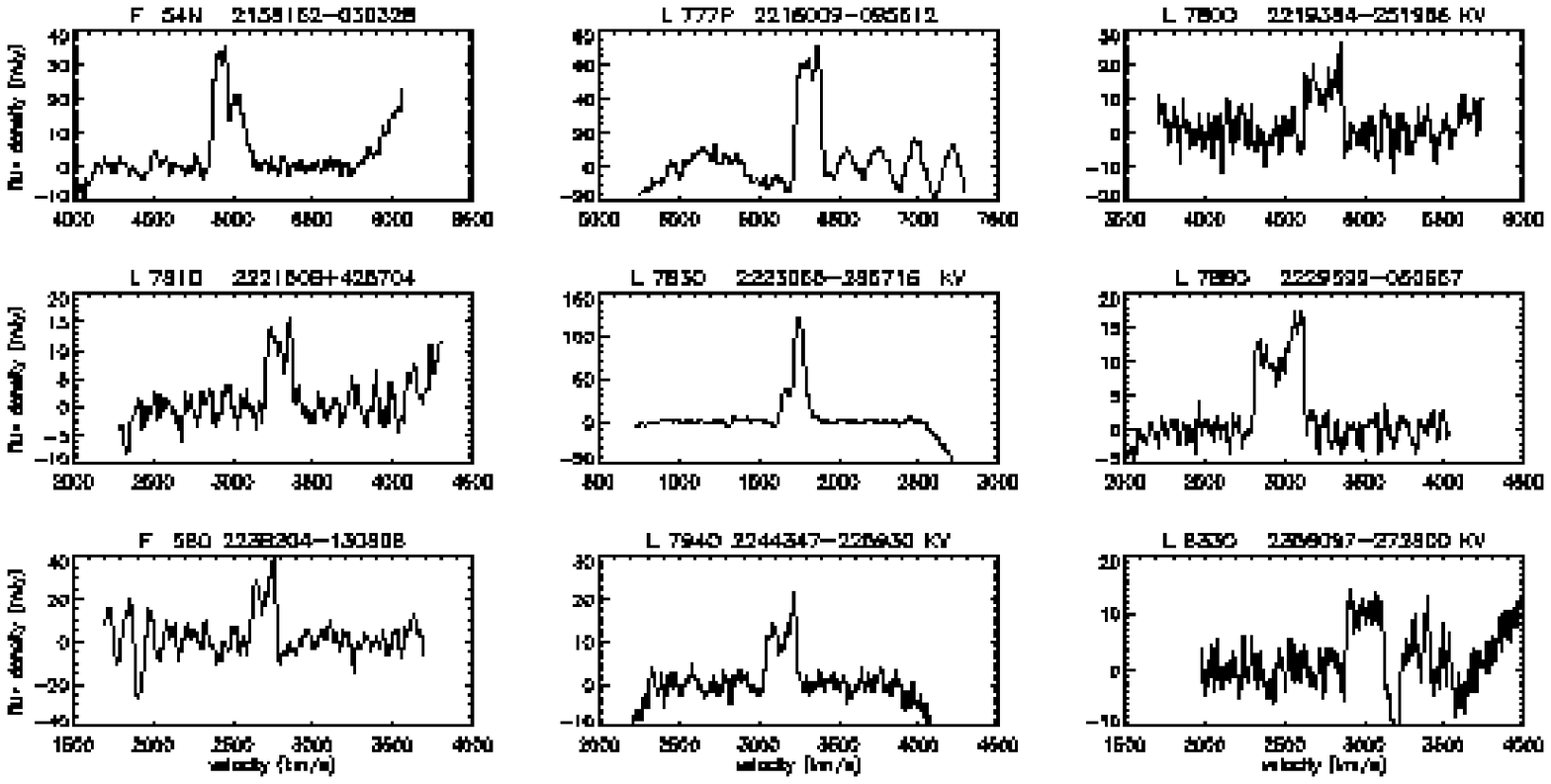}
\caption [] {{\bf h.}\, Nan\c{c}ay 21-cm \HI\ line spectra
of detected objects {\it - continued}. 
}
\end{figure*}

\begin{figure*} --- 3a -----------------
\centering
\includegraphics{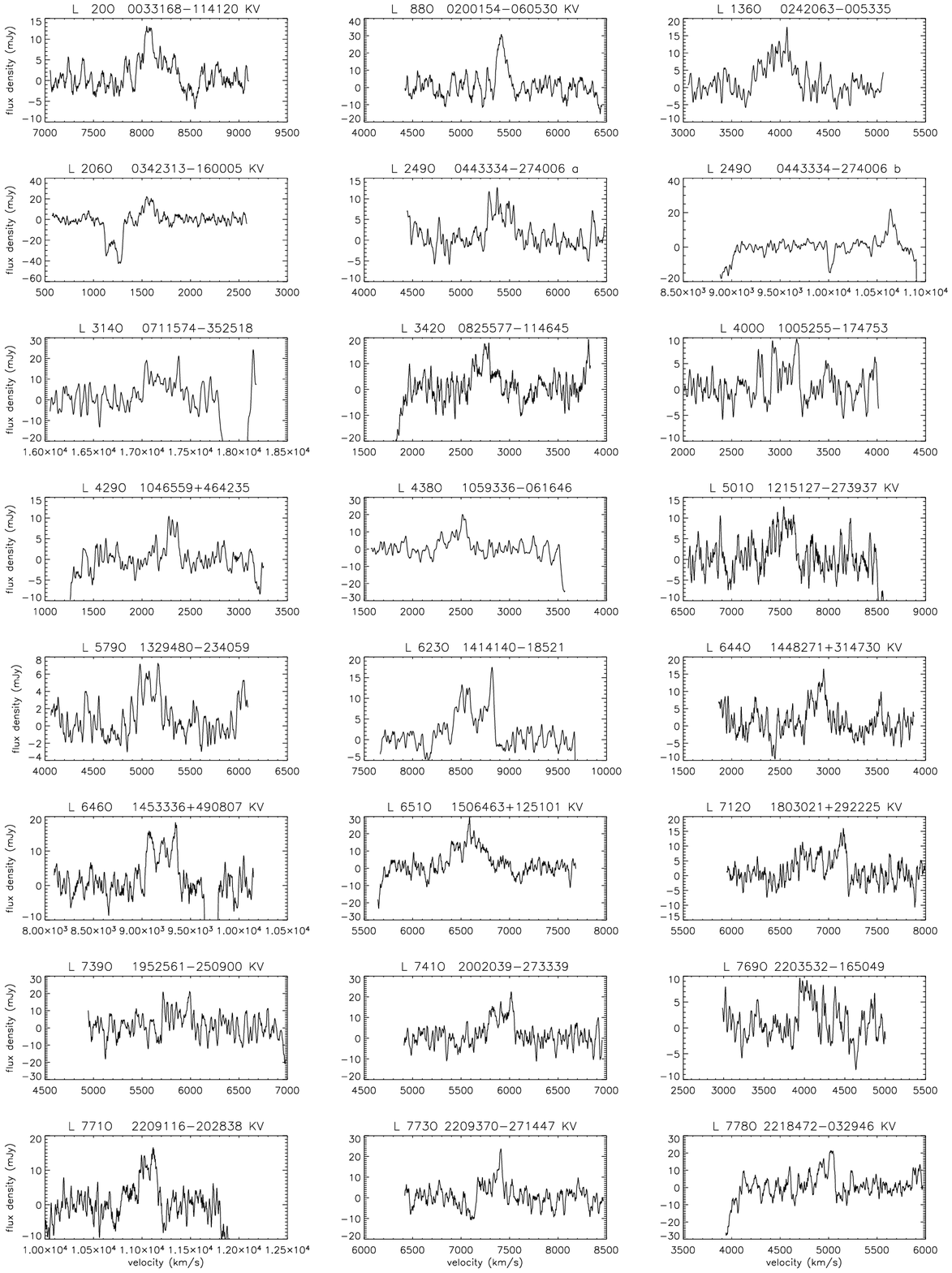}
\caption [] {{\bf a.}\, Nan\c{c}ay 21-cm \HI\ line spectra of marginal detections (see Table 4). 
Velocity resolution is 15.8 \kms\ (velocity search mode) and 17.1 \kms\ (known velocity mode), 
radial heliocentric velocities are according to the optical convention. Galaxies detected in the 
`known velocity' mode are indicated by the designation `KV' following their coordinates in the 
header of their spectrum.
}
\end{figure*}

  % Fig. 1 (detections) + Fig. 2 (marginal cases)

\begin{figure*} % --- Fig. 3 -----------------
\includegraphics[width=13cm,angle=90]{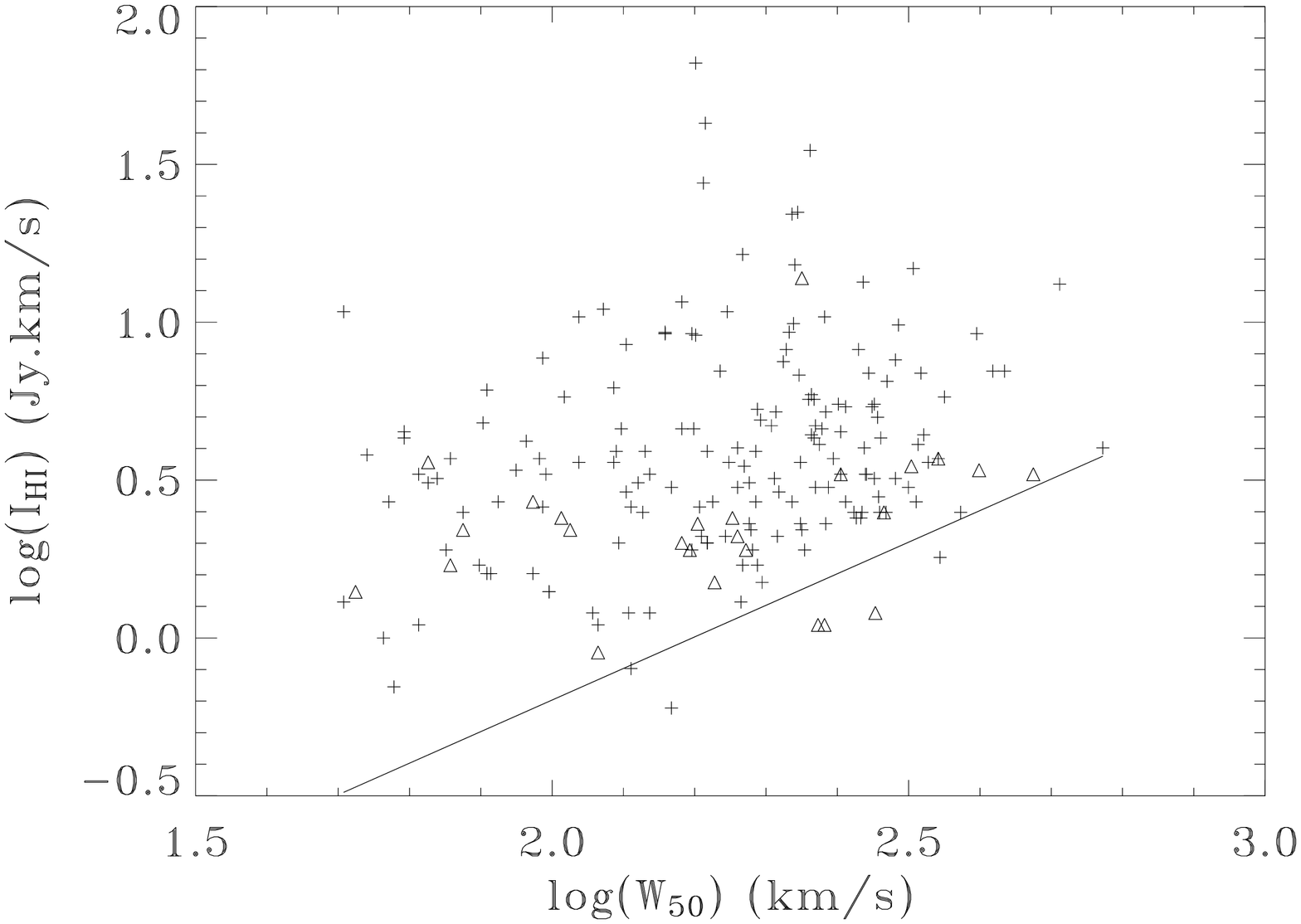}
\caption [] {Distribution of integrated \HI\ line fluxes $I_{HI}$ as a function of the \HI\ line FWHM, $W_{50}$.
The straight line indicates the 3 $\sigma$ detection limit for a 250 \kms\ wide, 
flat-topped spectral line, based on the average rms noise level of the data. 
Clear detections of survey galaxies are represented by crosses, and marginal detections by triangles.
}
\end{figure*}

\begin{figure*} % --- Fig. 4 -----------------
\includegraphics[width=13cm]{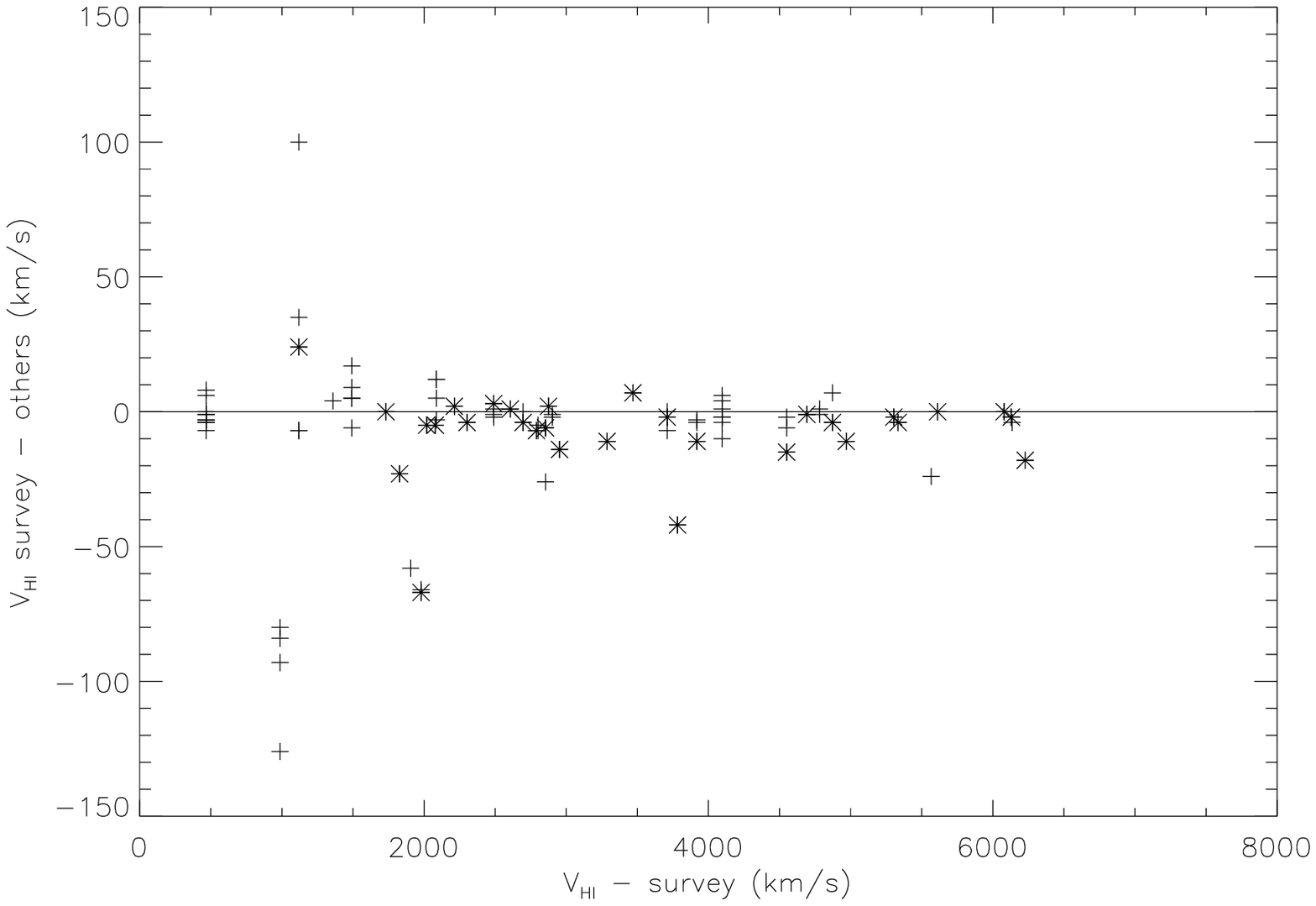}
\caption [] {Comparison of central  \HI\ profile velocities, $V_{HI}$, of survey and calibration galaxies with literature 
values (see Table 2). Plotted is the difference (survey-others) as function of the \HI\ radial velocity measured 
in the survey.
To guide the eye, a horizontal line was plotted at $\Delta$V = 0 \kms; this line does not represent a fit to the data.
Comparisons with published \nan\ data are represented by a $\star$, the crosses
indicate comparisons to measurements made with other telescopes.
}
\end{figure*}

\begin{figure*} % --- Fig. 5 -----------------
\includegraphics[width=13cm]{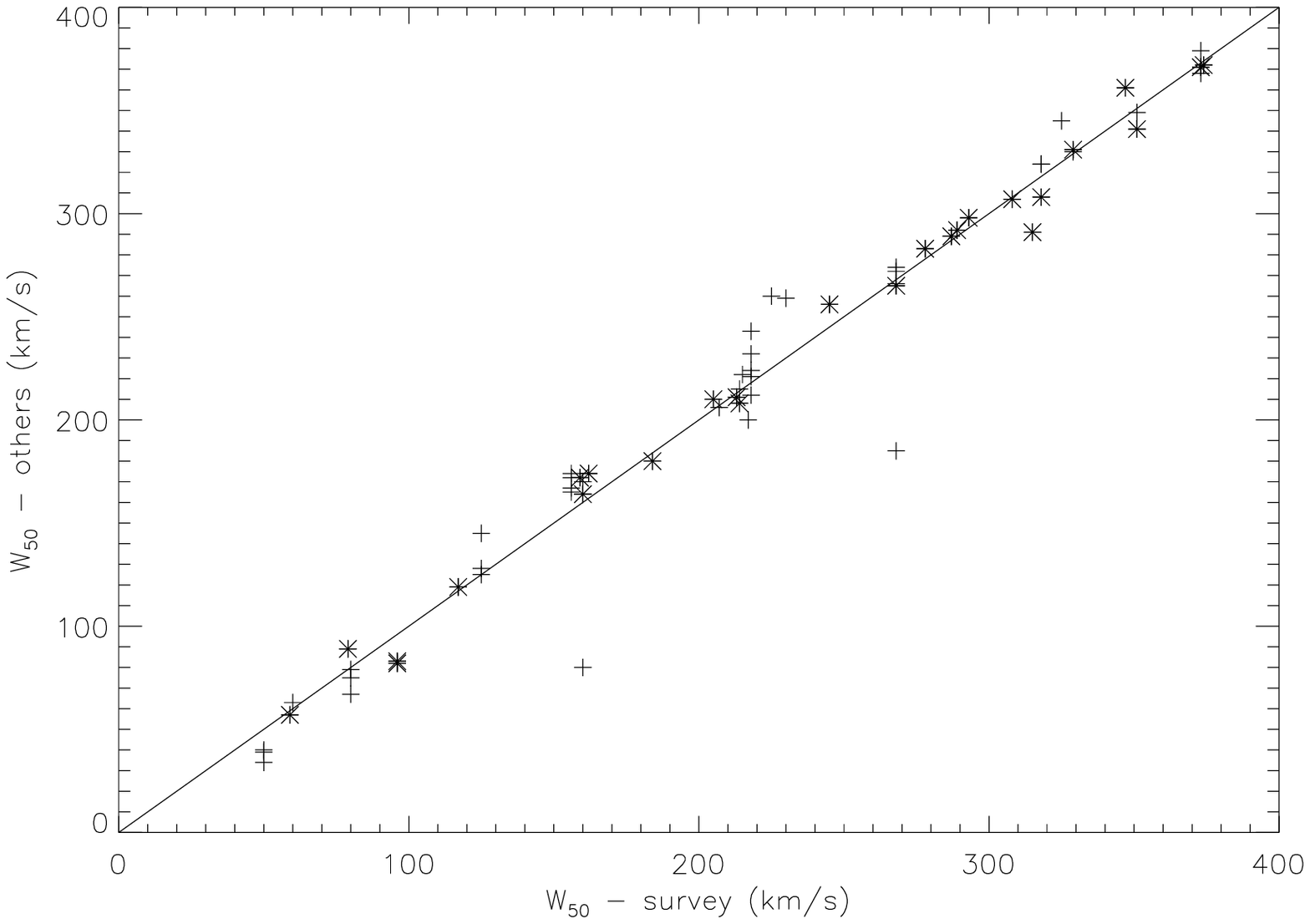}
\caption [] {Comparison of \HI\ profile FWHMs, $W_{50}$, of survey and calibration galaxies with literature values (see Table 2). 
To guide the eye, a diagonal line with a slope of 1 was plotted; this line does not represent a fit to the data.
Comparisons with published \nan\ data are represented by a $\star$, the crosses
indicate comparisons to measurements made with other telescopes.
}
\end{figure*}

\begin{figure*} % --- Fig. 6 -----------------
\includegraphics[width=13cm]{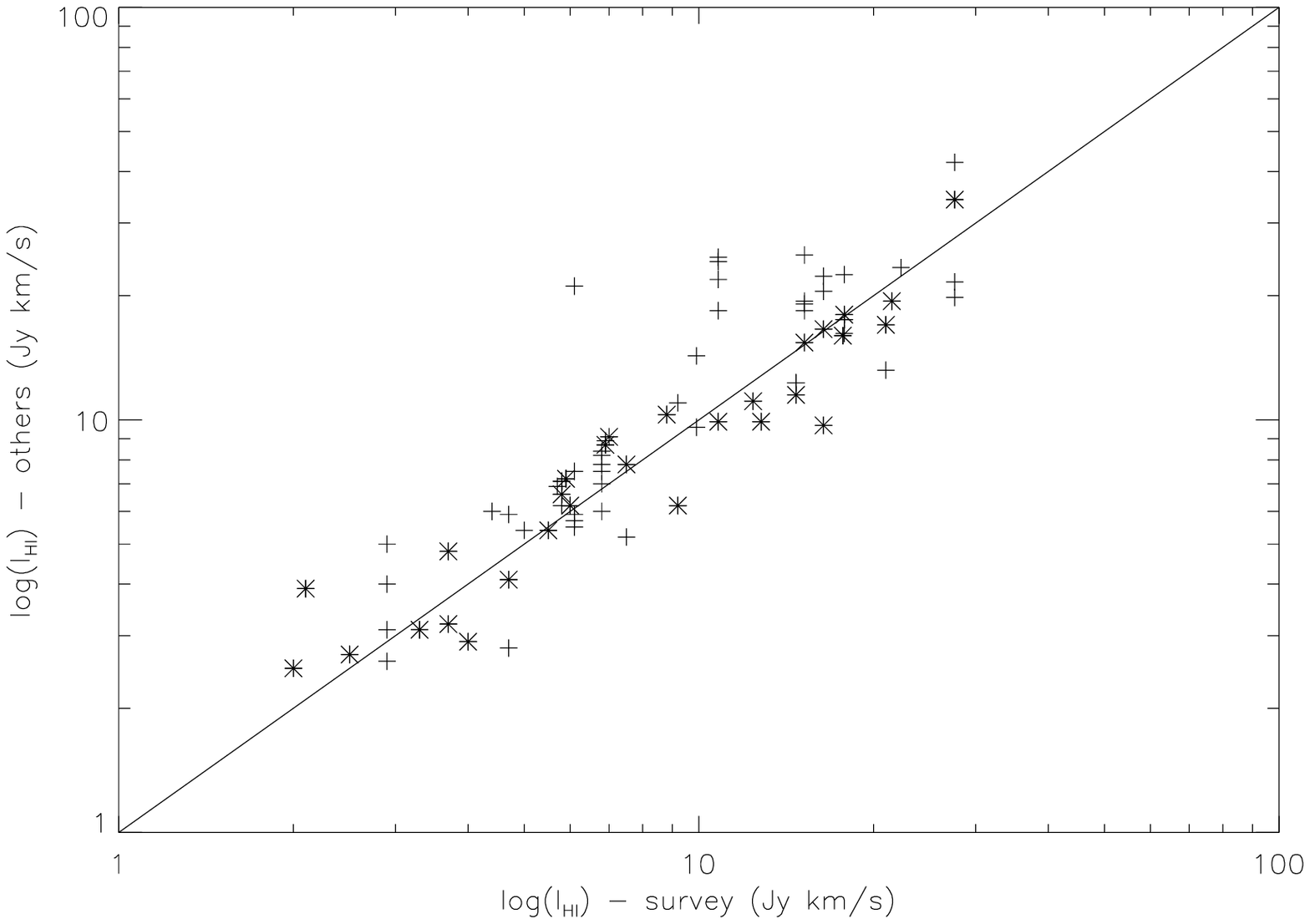}
\caption [] {Comparison of  integrated \HI\ line fluxes, $I_{HI}$, of survey and calibration galaxies with literature values 
(see Table 2). To guide the eye, a diagonal line with a slope of 1 was plotted; 
this line does not represent a fit to the data.
Comparisons with published \nan\ data are represented by a $\star$, the crosses
indicate comparisons to measurements made with other telescopes.
}
\end{figure*}

\end{document}